\documentclass[useAMS,usenatbib]{mn2e}

\usepackage{graphicx}
\usepackage{latexsym}
\usepackage{amsfonts,amsmath,amssymb}
\usepackage{url}
\usepackage[utf8]{inputenc}
\usepackage{fancyref}
\usepackage{hyperref}
\hypersetup{colorlinks=true,pdfborder={0 0 0}, citecolor= blue}
\usepackage{natbib}
\usepackage{multirow}

\newcommand{\oii}{[O~{\sc ii}]}
\newcommand{\oiii}{[O~{\sc iii}]}
\newcommand{\neiii}{[Ne~{\sc iii}]}
\newcommand{\nev}{[Ne~{\sc v}]}
\newcommand{\feii}{Fe~{\sc ii}}
\newcommand{\hb}{H$\beta$}
\newcommand{\kms}{${\rm km~s^{-1}}$}

\newcommand{\aj}{ApJ}
\newcommand{\apj}{AJ}
\newcommand{\apjl}{ApJL}
\newcommand{\apjs}{ApJS}
\newcommand{\araa}{ARA\&A}
\newcommand{\aap}{A\&Ap}
\newcommand{\nat}{Nature}
\newcommand{\mnras}{MNRAS}
\newcommand{\pasp}{PASP}

\begin{document}

\title{Tracing Quasar Narrow-Line Regions Across Redshift:\\ A Library of High
S/N Optical Spectra}

\author[Tammour, Gallagher, and Richards]{A. Tammour\thanks{atammour@uwo.ca}$^1$, S. C. Gallagher$^1$, \& Gordon Richards$^2$ \\ $^1$ University of Western Ontario \\ $^2$ Drexel University}
\maketitle

\begin{abstract}
In a single optical spectrum, the quasar narrow-line region (NLR) reveals low density, photoionized gas in the host galaxy interstellar medium, while the immediate vicinity of the central engine generates the accretion disk continuum and broad emission lines.
To isolate these two components, we construct a library of high S/N optical composite spectra created from the Sloan Digital Sky Survey (SDSS-DR7).  
We divide the sample into bins of continuum luminosity and \hb\ FWHM that are used to construct median composites at different redshift steps up to 0.75.
We measure the luminosities of the narrow-emission lines \nev$\lambda$3427, \neiii$\lambda$3870, \oiii$\lambda$5007, and \oii$\lambda$3728 with ionization potentials (IPs) of 97, 40, 35, and 13.6 eV respectively.
The high IP lines' luminosities show no evidence of increase with redshift consistent with no evolution in the AGN SED or the host galaxy ISM illuminated by the continuum.
In contrast, we find that the \oii\ line becomes stronger at higher redshifts, and we interpret this as a consequence of enhanced star formation contributing to the \oii\ emission in host galaxies at higher redshifts.  
The SFRs estimated from the \oii\ luminosities show a flatter increase with $z$ than non-AGN galaxies given our assumed AGN contribution to the \oii\ luminosity.
Finally, we confirm an inverse correlation between the strength of the \feii$\lambda4570$ complex and both the \oiii\ EW (though not the luminosity) and the width of the \hb\ line as known from the eigenvector 1 correlations. 
\end{abstract}


\section{Introduction}

The current paradigm of active galactic nuclei (AGN) includes a central source that comprises a super-massive black hole (SMBH) surrounded by an accretion disk \citep{antonucci93, urry95}.  
Through this accretion disk, material gets funnelled into the SMBH \citep{shakura73}.  
Quasars are the most luminous class of AGN with bolometric luminosities of up to $10^{48}~{\rm erg~ s^{-1}}$.

Photoionization by radiation from the accretion disk is thought to be the main mechanism through which gas in AGNs is ionized \citep{osterbrock06}.  
Studying the interactions between the ionizing photons and the material they encounter allows us to study the physical conditions in the ionized regions.  
For example, the presence of narrow forbidden lines in quasar spectra points to a region that has low density and is located relatively far from the black hole's sphere of influence.  
For the most part, the gas in the NLR region responds kinematically to the gravitational potential of the galaxy as a whole and therefore produces narrower emission lines (FWHM $\le 1000$ \kms) -- hence its label.  
On the other hand, in the close vicinity of the central source ($\leq$ 1 pc), the gas mainly feels the gravitational influence of the SMBH and therefore produces emission lines that are primarily Doppler-broadened with widths of 1000s km/s. 
We therefore call this region the broad-line region (BLR).

Though the broad and narrow-line regions are physically distinct, correlations between properties of the two regions are found in the literature.  
Most well-known are the so-called eigenvector 1 relations, first identified by \citet[][hereafter BG92]{boroson92} from principal component analysis of several spectral measurements of the low-redshift PG quasar sample.  
BG92 found that narrow \hb\ width, strong \feii\ emission, and weak \oiii\ emission tended to be found in the same objects.  
The strong eigenvector 1 objects have also been identified as having relatively high $L/L_{\rm Edd}$ values \citep[e.g.,][]{boroson92,boroson02,sulentic07}.  
Given our understanding of the rather different origins of BLR and NLR emission, the linking of broad- and narrow-line spectral properties is puzzling, and the extended discussion in BG92 reflects this conceptual difficulty.

To first order, the NLR can be described as the diffuse interstellar medium of the host galaxy illuminated by the accretion disk continuum.
Quasar NLRs are therefore the regions where we expect the interactions between AGNs and their host galaxies to take place; the narrow emission lines are among the few tracers of the host galaxy that are evident in optical quasar spectra.  
It is therefore important to check that any apparent links between broad and narrow-line region spectral properties do not arise from secondary and unintentional selection effects of host galaxies.  
For example, the strong eigenvector 1 objects are necessarily low redshift (because of the lines used in the study), and therefore low luminosity.  
The original BG92 sample were all PG quasars, which because of their selection are not representative of the quasar population as a whole \citep{jester05}.

Therefore, we are interested in studying samples of quasars with very similar central engines across a range of redshift to determine if we can isolate the effects of host galaxy evolution on NLR emission.
With the large Quasar Properties Catalogue of \citet{shen11} with 105,783 objects, this hypothesis can be tested by selecting similar types of quasars (in terms of $L_{\rm opt}$ and $M_{\rm BH}$), and examining their narrow-line region properties independently.

Throughout this work we use: $\Omega_{\Lambda} = 0.7,~ \Omega_0 =0.3,~h=0.7$ \citep{spergel03}.

\section{Data and Analysis}

\subsection{The Sample}

The initial sample is selected from the \citet{shen11} catalogue of quasar properties \citep[SDSS-DR7;][]{schneider10,shen11}.  
We restrict the sample to radio-quiet quasars (defined in the catalogue as objects with $R=f_{6 cm}/f_{2500} < 10$; where $f_{6 cm}$ and $f_{2500}$ are the radio and optical flux densities observed at 6 cm and 2500 \AA\ respectively).  
We also impose a redshift limit of $z \leq 0.75$ to ensure $\lambda_{rest}=5100$ \AA\ is in our spectral range.  
This emission-line-free segment of the continuum is used to first normalize the spectra and later to calculate the relative line luminosities of the composites.  
The final sample contains $16,027$ spectra.  
We utilize the OH-subtracted spectra of \citet{wildoh10} who constructed a set of the telluric-line-cleaned spectra (specifically around \hb).  
We then apply IRAF's DEREDDEN tool\footnote{IRAF is distributed by the National Optical Astronomy Observatories, which are operated by the Association of Universities for Research in Astronomy, Inc., under cooperative agreement with the National Science Foundation.} to correct the spectra for Galactic extinction using the $E(B-V)$ values provided in the catalogue \citep [][and references therein]{shen11}.  
Finally, the spectra are shifted to the rest frame using the \citet{hewett10} redshift determination as provided in \citet{shen11}.

\subsection{The Composites}
\label{composites}

One of the advantages of starting from a large homogeneous dataset like the SDSS quasar sample is the ability to stack objects to create higher signal-to-noise spectra which then allows us to examine weaker features (such as the high IP \nev\ line) that are otherwise hard to measure \citep[e.g,][]{vandenberk01, sulentic02, croom02, hill14}.  
To make sure we are identifying objects with similar AGN properties, we group objects according to their continuum luminosities at 5100 \AA\ ($L_{5100}$) and the FWHM of \hb\footnote{Hereafter when we refer to \hb, we mean the broad component of the line from which a narrow-line component has been subtracted.} as reported in \citet{shen11}.  
These two quantities are often used in combination in the literature to estimate more physical properties of the AGN such as the mass of the SMBH and the accretion rate \citep[][and references therein, see also \S \ref{measurements}]{shen11}.  
We used the measured properties (luminosity and Balmer-line broad-line width) rather than derived properties ($M_{\rm BH}$ and $L/L_{\rm Edd}$) to keep the binning process closer to the data.  
As shown later, in practice, our choice of binning also means that objects with similar values of $M_{\rm BH}$ and $L/L_{\rm Edd}$ are grouped together.  
One caveat is that the conversion of \hb\ FWHM and $L_{5100}$ into BH masses and $L/L_{Edd}$ can be influenced by several factors such as the dependence of FWHM on our inclination angle to the broad-line region and the choice of bolometric correction for $L_{5100}$  \citep[e.g,][]{richards06}.
We start by sorting the objects according to their $L_{5100}$ values and then dividing them into 10 groups with an equal number of objects (1602 or 1603).  
We choose to look at four of these groups with the low ($L1$: $\log L_{5100} = 43.15 - 44.17$ erg s$^{-1}$), intermediate ($L4$: $\log L_{5100} = 44.36 - 44.45$ erg s$^{-1}$ and $L7$: $\log L_{5100}=44.63 - 44.72$ erg s$^{-1}$), and high ($L10$: $\log L_{5100} = 44.98 - 46.19$ erg s$^{-1}$) luminosities.  
Specifically, we used groups 1, 4, 7, and 10 and skipped the intermediate ones (2, 3, 5, 6, 8, and 9).
This choice gives us the full dynamic range of the sample, while keeping the number of bins to a manageable number.  

We do not increase the size of the bins to use all of the spectra in the sample because we are interested in creating high S/N composites of objects that are as similar as possible.  
The conversion of $L_{5100}$ and \hb\ FWHM into $L/L_{\rm Edd}$ and $M_{\rm BH}$ has significant uncertainties \citep[e.g,][]{vestergaard06, shen11}.  
For example, high energy AGN spectral energy distributions change as a function of luminosity \citep[e.g.,][]{steffen06}, and so derived quantities such as bolometric luminosities (required for $L/L_{Edd}$) are likely not properly calibrated for all luminosities \citep{richards06bol}.

We then use the measurements of \hb\ from \citet{shen11} to separate objects into three groups: narrow, intermediate, and broad \hb\ using 2000 and 4000 \kms\ as our boundaries (excluding objects with \hb\ FWHM $>$ 20000 \kms).  
We choose fewer bins in \hb\ FWHM compared to $L_{5100}$ because of the considerably smaller dynamic range of the FWHM values, a factor of $\approx20$ compared to three orders of magnitude for the sample's range of $L_{5100}$.  
The boundary at 2000~\kms\ marks the traditional definition of narrow-line Seyfert~1s \citep{osterbrock85}, and the intermediate bin covers the main peak of
the \hb\ FWHM distribution.

Once the quasars are divided by luminosity and Balmer line width, we are interested in looking at the NLR properties of these bins.
Specifically, we investigate if there is any evidence for redshift evolution that might be expected if AGN host galaxies have evolved since redshift 0.75.  
We thus divide each of these luminosity groups into 7 subsets according to their redshifts with bins of width 0.1 centred at: 0.1, 0.2, 0.3, 0.4, 0.5, 0.6, and 0.7.  
This results in 4 $L_{\rm 5100}$ $\times$ 3 \hb\ FWHM $\times$ 7 $z$ = 84 bins, 8 of which are empty (primarily the high luminosity/low redshift ones) as shown in Table \ref{tbl:bins}.  
For the 76 non-empty bins, we make median composite spectra with $3 \sigma$ clipping using IRAF's SCOMBINE tool.  
We choose to create median composites to preserve the EWs of the lines (see \citet[][]{vandenberk01} for a good discussion of the difference between median and mean composites).

\begin{table*}
\caption{Breakdown of the number of objects in each luminosity subset with narrow (n; FWHM $<$ 2000 \kms), intermediate (i; 2000 \kms $<$ FWHM $<$ 4000 \kms), and broad (b; 4000 \kms $<$ FWHM $<$20,000 \kms) \hb\ at each redshift step. We only use the bins with $> 10$ objects to create composites. Those bins are highlighted with boldface in this table.}
\label{tbl:bins}
\begin{tabular}{cccc c ccc c ccc c ccc}
\hline
$\log~L_{5100}$ & \multicolumn{3}{c}{$L1$} && \multicolumn{3}{c}{$L4$} && \multicolumn{3}{c}{$L7$} && \multicolumn{3}{c}{$L10$} \\
$[{\rm erg~s}^{-1}]$& \multicolumn{3}{c}{43.15 -- 44.17} && \multicolumn{3}{c}{44.36 -- 44.45} && \multicolumn{3}{c}{44.63 -- 44.72} && \multicolumn{3}{c}{44.98 -- 46.19} \\
\hline
Redshift &n      &i   &b  &  &n    &i     &b  & &n      &i     &b & &n     &i  &b \\ 
\hline
0.1    & 14   &  29      & 29 & 	    &   {\bf 0} 	 &  {\bf 3} 	        &  {\bf 1} 	 &     &   {\bf 0} 	   &   {\bf 0} 	 &  {\bf 1} 	  &      &  {\bf 0}       &   {\bf 0} 	& {\bf 0} \\
0.2 	&  55    & 203     & 249&       & {\bf 10}       &  20 	&  23  &	&   {\bf 1} 	   &{\bf 7} 	          & {\bf 9} 	    &    & {\bf 0} 	      &   {\bf 0}    & {\bf 1} \\
0.3 	&  40    & 168     & 255 &      &  22 	 &  92 	& 126 &	&   {\bf 6} 	   & 36 	 &  28    &   &   {\bf 2}      &  19 	&  {\bf 10} \\
0.4 	&  34    &  75      &  91   &     & 56 	 & 238 	& 299 &	&  27   & 81 	 & 85     &   &  16     &  37 	&  49 \\
0.5 	&  12    &  72      &  68   &     &  34 	 & 141 	& 188 &	&  19    &143 	 &216    &   &   {\bf 8}      & 94 	& 108 \\
0.6 	&  13    &  38      &  82    &     &  {\bf 8}      & 85 	& 104   &  &  37   & 252 	 & 306   &   &27       &178 	& 232 \\
0.7 	&  {\bf 6}      & 21 	     &  41  &   &  14 	 &  51     & 84 	   &   & 23 	   & 137     & 184   &   & 30      &314 	& 475 \\
\hline
Total & 174 	& 606        &815 &       & 144        & 630       & 825  &     & 113   &656 	 &830  & &  83   & 642       &875 \\
\hline
   \end{tabular}
\end{table*}

\begin{figure}
\begin{center}
\includegraphics[width=1\columnwidth]{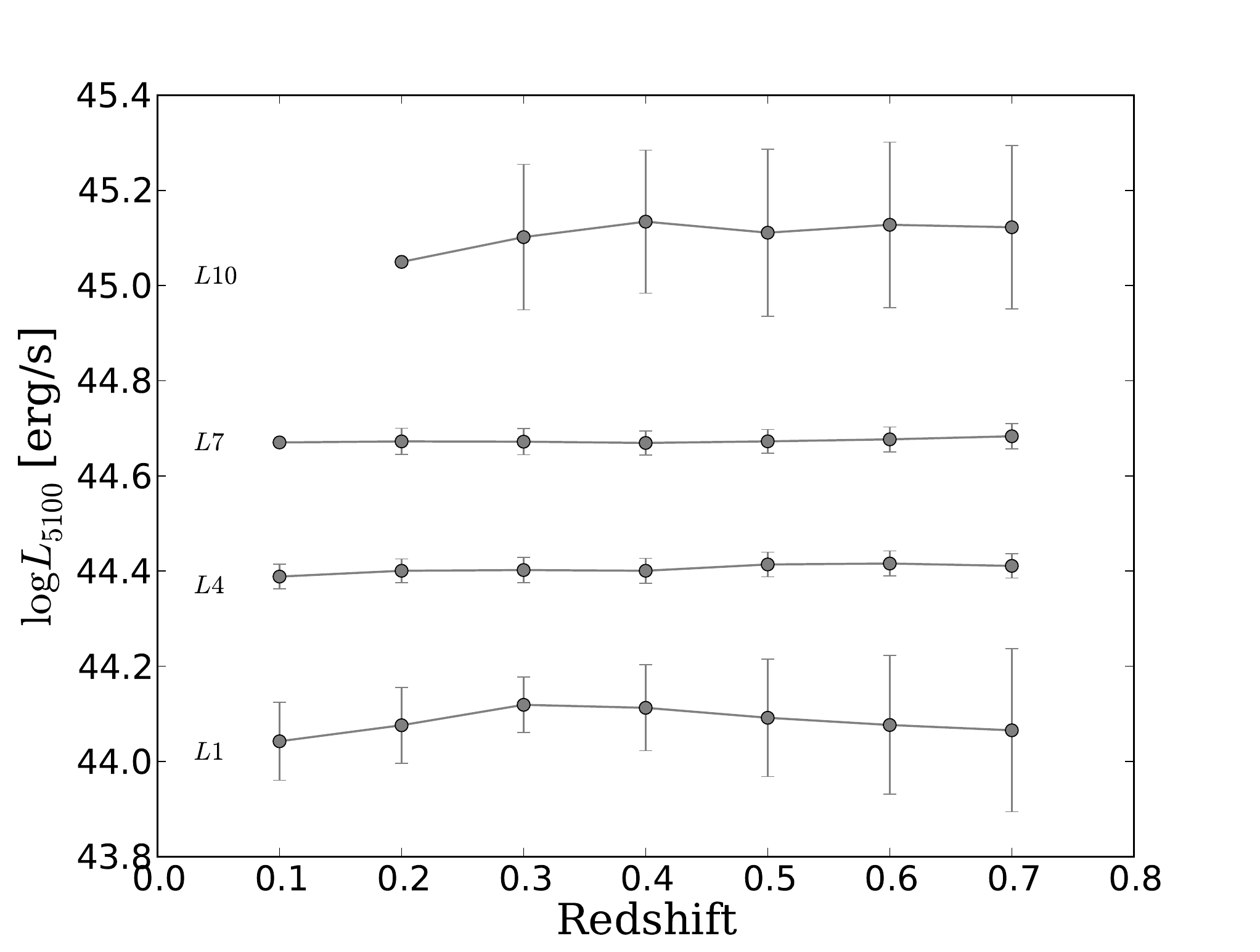}
\caption{Median $L_{5100}$ from \citet{shen11} vs. redshift. Within the given uncertainty, the continuum luminosity is constant within the same bin across redshift. The $L10$ group does not have objects with redshift $< 0.2$. The error bars show the standard deviation in the bin. The two data points without error bars represent bins with fewer than 3 objects.}
\label{fig:l5100vsz} 
\end{center}
\end{figure}

\begin{figure}
\begin{center}
\includegraphics[width=1\columnwidth]{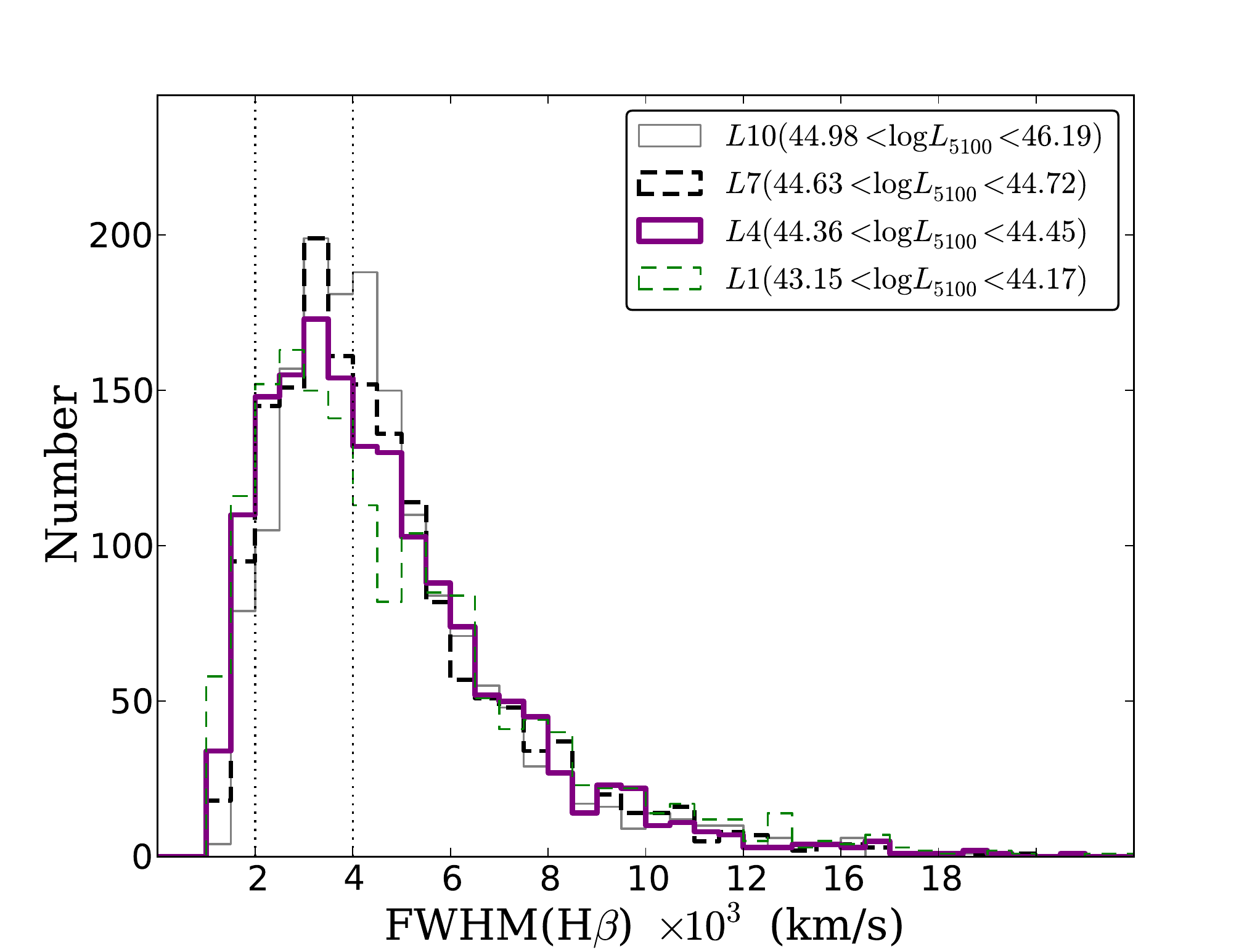}
\caption{Distribution of \hb\ FWHM for the four luminosity subsets. The distributions look similar except for a weak trend of rising numbers of broader \hb\ objects in higher continuum luminosity bins (grey solid line). Narrower \hb\ quasars are more abundant in the lowest luminosity bin (green dashed line). Objects with \hb\ FWHM $>$ 20,000 \kms\ are not included in making the composites as these were often found to arise from either bad measurements or low quality spectra.}
\label{fig:fwhmhisto}
\end{center}
\end{figure}

\begin{figure}
\begin{center}
\includegraphics[width=1\columnwidth]{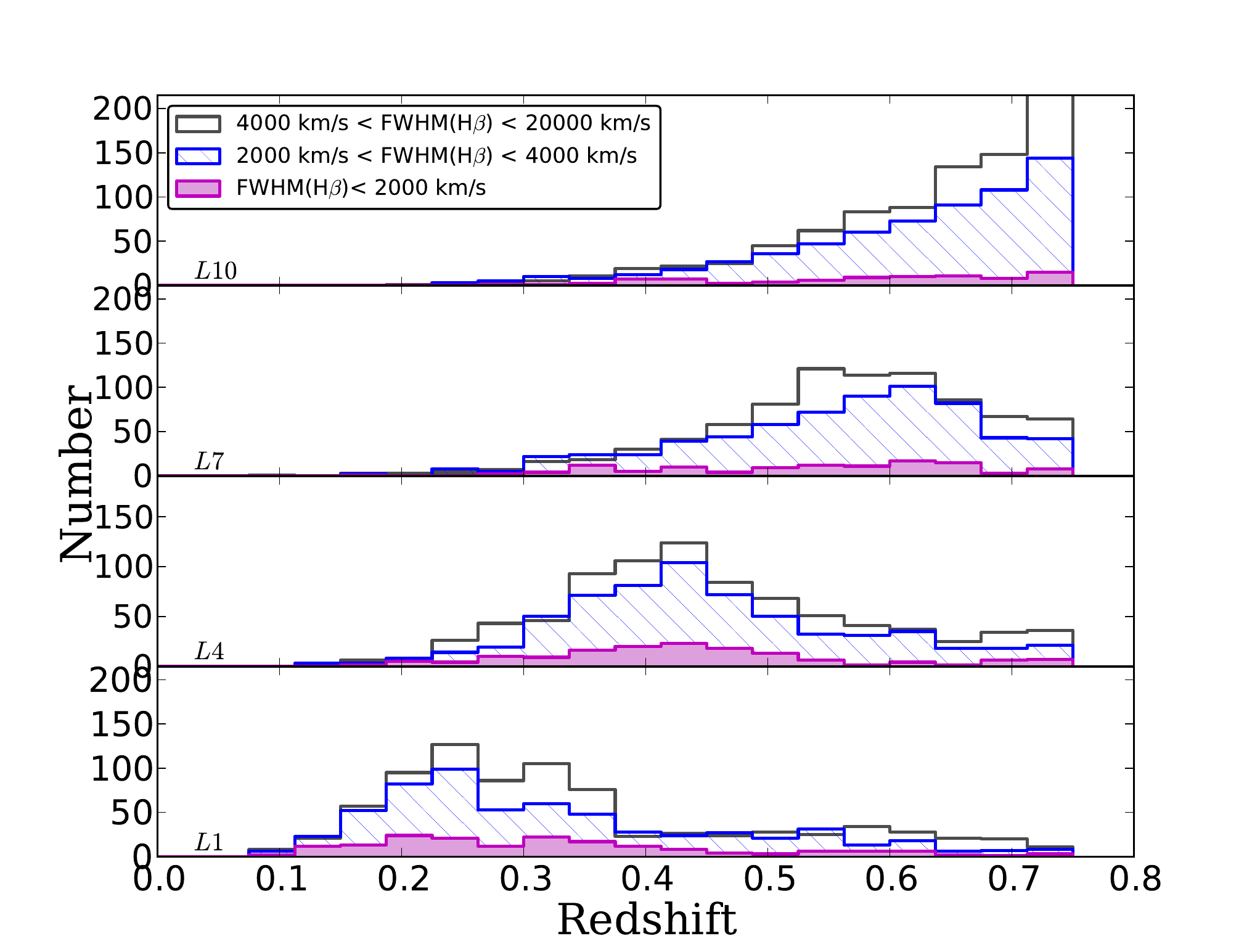}
\caption{Distribution of redshift at the four continuum luminosity levels for the narrow (solid magenta), intermediate (hashed blue) and broad (open grey) \hb\ subsets. There is a gradual shift in the distributions of redshift from lower luminosity to the higher luminosity subsets -- higher luminosity objects have more objects at higher redshift than lower luminosity ones as expected from the SDSS flux limit for quasar spectroscopy and the redshift evolution of the quasar luminosity function.}
\label{fig:zhisto} 
\end{center}
\end{figure}

To examine the distributions of objects within our binning, we look at the distributions of $L_{5100}$, \hb\ FWHM, and redshift (Fig. \ref{fig:l5100vsz}, \ref{fig:fwhmhisto}, and \ref{fig:zhisto}).
The figures show that the distributions are fairly consistent among the different bins.  
We note, however, that there are more quasars with higher luminosity at higher redshift as expected because of the SDSS flux limit for quasar spectroscopy ($i < 19.1$) and the redshift evolution of the quasar luminosity function \citep{richards06}.  
Table \ref{tbl:bins} also shows that the narrow \hb\ subsets contain fewer objects than the intermediate and broad subsets.  
This is mainly due to the selection of objects in the SDSS quasar database requiring at least one broad line with FWHM $> 1000$ \kms\ \citep{schneider10}.  
The spectral library is given in Appendix \ref{compositelib} (available in the online-only version).

\subsection{Measurements}
\label{measurements}
\begin{figure*}
\includegraphics[width=1.5\columnwidth]{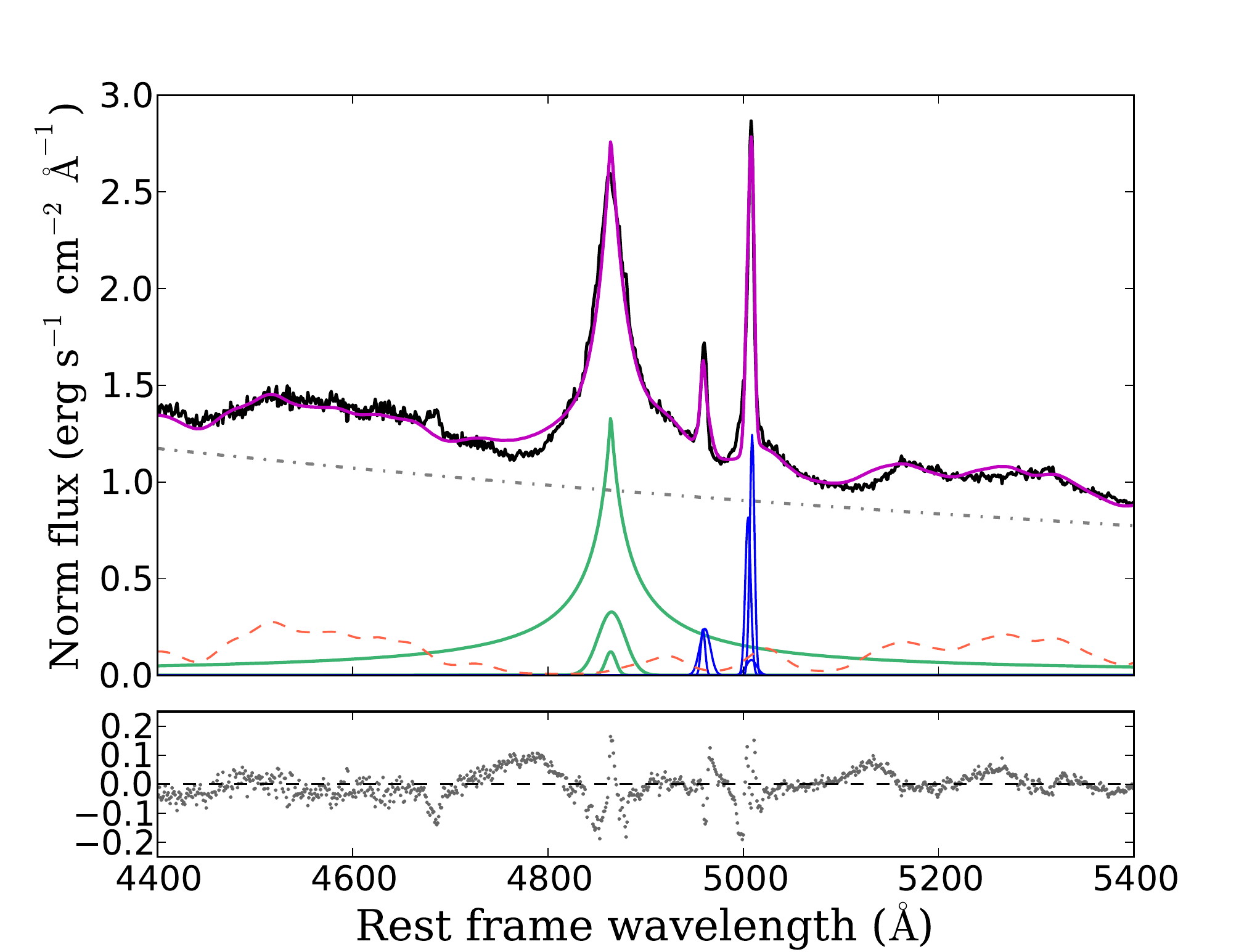}
\caption{Model fit for composite iL7z2 (black) and the full fitting result (magenta). \hb\ is fit with a Lorentzian and up to two Gaussians (green), the \oiii\ doublet is fit with up to three Gaussians per line (blue), the grey dash-dot line is a power-law continuum, and the \feii\ template is shown as a dashed red line. The residuals are shown in the lower panel.}
\label{example_fit}
\end{figure*}

Optical quasar spectra show several identifiable narrow emission lines.  
For this study, we focus on four narrow emission lines that are available in the wavelength range of our composites: \nev\ $\lambda3427$, \oii\ $\lambda3728$, \neiii\ $\lambda3870$, and \oiii\ $\lambda5007$, with ionization potentials of 97.1, 13.6, 40, 35 eV, respectively \citep{peterson97}.
These are the brightest, common emission lines, and they span a large range of ionization potentials.  
The high S/N present in most of our composites allows us to measure some of the relatively weak features such as \neiii\ $\lambda3728$ that are typically not detected in individual spectra.
We measure the flux of these narrow lines in each composite using one of two approaches. 
First, in the \hb\ - \oiii\ $\lambda\lambda 4959,5007$ region, we use IRAF's SPECFIT package to fit the composite with a multicomponent model \citep{kriss94}. 
This method allows us to subtract the \feii\ emission features which form a pseudo-continuum of blended lines that are in some cases heavily blended with other emission lines in this part of the spectrum. 
For the removal of the \feii\ lines, we make use of the \feii\ template of \citet{veron-cetty04}.
We use a Lorentzian and up to two Gaussians to fit \hb, and up to three Gaussians to fit each of the \oiii\ lines.
An example of the fitting model in this spectral region is shown in Figure \ref{example_fit}.
In the \neiii\ - \nev\ - \oii\ ($\sim 3300 - 4000$ \AA\ ) region we use IRAF's SPLOT tool and simply integrate the flux above the continuum (defined around the emission line by visual inspection) as we do not anticipate significant contamination from the \feii\ lines.

The results of the measurements are tabulated in Table~\ref{big_table}. 

We use the $L_{5100}$ and \hb\ width measurements to calculate the BH mass and accretion rate for each composite.
For the BH masses, we use equation 5 from \citet{vestergaard06}:
\begin{equation}
\begin{aligned}
\log \frac{M_{BH}}{M_{\odot}} = 0.910 + 0.5 \log  \left ( \frac{5100L_{5100}}{10^{44} \rm{erg \ s^{-1}}} \right ) \\
+ 2 \log \left ( \frac{\rm{FWHM (H}\beta)}{\rm{km \ s^{-1}}} \right ) .
\end{aligned}
\end{equation}
For the accretion rates, we estimate the bolometric luminosity as $L_{\rm bol} = 9 L_{5100}$ \citep{peterson97}.
The Eddington luminosity, $L_{\rm{Edd}}$, is calculated as \citep{netzer06}:
\begin{equation}
L_{\rm{Edd}} \simeq 1.5 \times 10^{38} \left ( \frac{M_{BH}}{M_{\odot}} \right ) \rm{erg \ s^{-1}}.
\end{equation}

Values of the calculated $M_{BH}$, $L_{\rm{Edd}}$, and $L_{\rm bol}$ for each composite are given in Table \ref{tbl:mbh}.  
\citet{richards06bol} have expressed some reservations about the potential pitfalls of applying a single bolometric correction to all quasars as it is quite likely that different quasar populations should have distinct bolometric correction values because of systematic differences in their spectral energy distributions \citep{krawczyk13}.  
This concern was among the reasons that the samples comprising each composite are binned by $L_{5100}$ and H$\beta$ width alone rather than the derived quantities of $M_{BH}$ and $L/L_{Edd}$.  
We emphasize that these reported values of $M_{\rm BH}$, $L_{\rm{Edd}}$, and $L_{\rm bol}$ are not used for the binning of objects for the composite selection, but are simply reported for completeness.  

\begin{table}
\caption{BH masses, $L_{\rm bol}$ and $L_{\rm{Edd}}$ for the composites in each of the $L_{5100}$ and H$\beta$ width bins.}
\label{tbl:mbh}
\begin{tabular}{ccccc}
\hline
Composite & $\log (M_{BH}/M_{\odot}$) & $ \log L_{\rm{Edd}}$ & $\log L_{\rm bol}$  & L/L$_{\rm Edd}$\\
& & [erg s$^{-1}$] & [erg s$^{-1}$] & \\
\hline
nL1 & 7.39 & 45.57 & 45.06 & 0.31 \\
iL1 & 7.90 & 46.08 & 45.05 & 0.09 \\
bL1 & 8.54 & 46.71 & 45.05 & 0.02 \\
nL4 & 7.57 & 45.75 & 45.37 & 0.41 \\
iL4 & 8.00 & 46.18 & 45.36 & 0.15 \\
bL4 & 8.63 & 46.81 & 45.36 & 0.04 \\
nL7 & 7.73 & 45.90 & 45.63 & 0.53 \\
iL7 & 8.22 & 46.40 & 45.63 & 0.17 \\
bL7 & 8.74 & 46.92 & 45.63 & 0.05 \\
nL10 & 7.98 & 46.16 & 46.06 & 0.80 \\
iL10 & 8.47 & 46.64 & 46.08 & 0.28 \\
bL10 & 8.94 & 47.12 & 46.07 & 0.09 \\
\hline
\end{tabular}
\end{table}

\section{Measured Line Luminosities}
\label{big_table}

The table below (Table \ref{tbl:linelum}) includes the measured line luminosities for the four narrow lines discussed in this paper (\nev, \neiii, \oii, and \oiii).
The full table is available online as an electronic table. 
\begin{table*}
\caption{Measured line luminosities for the four forbidden lines \oii, \oiii, \neiii, and \nev. The full table is available online as an electronic table. 
\label{tbl:linelum}}
\begin{tabular}{cc|cccc}
\hline
Composite & z & L(\nev)                              &  L(\neiii)                           & L(\oii)                                & L(\oiii)  \\
                   &    & $\times 10^{40} {\rm erg~s}^{-1}$ &$ \times 10^{40} {\rm erg~s}^{-1}$ & $\times 10^{40} {\rm erg~s}^{-1}$  &$\times 10^{40} {\rm erg~s}^{-1}$\\ 
\hline                  
nL1z1& 0.1 & 17.08 $\pm$ 0.79                 & 5.55 $\pm$ 0.51        & 7.29 $\pm$ 0.50        & 23.68 $\pm$ 5.97 \\
nL1z2 & 0.2 & 9.88 $\pm$ 0.41                & 6.68 $\pm$ 0.45        & 9.49 $\pm$ 0.39        & 21.71 $\pm$ 4.62 \\
nL1z3 & 0.3 & 10.18 $\pm$ 0.78                & 11.02 $\pm$ 0.77        & 12.25 $\pm$ 0.72        & 21.09 $\pm$ 1.14 \\
nL1z4 & 0.4 & 11.74 $\pm$ 0.96               & 13.11 $\pm$ 0.99        & 12.62 $\pm$ 0.81        & 25.56 $\pm$ 1.65 \\
nL1z5 & 0.5 & 7.95 $\pm$ 1.99                & 5.61 $\pm$ 1.70           & 14.93 $\pm$ 2.07         & 34.09 $\pm$ 2.07 \\
nL1z6 & 0.6 & 18.03 $\pm$ 5.41                & 15.21 $\pm$ 5.60           & 22.80 $\pm$ 4.55         & 72.31 $\pm$ 7.37 \\
nL1z7 & 0.7 & 28.98 $\pm$ 9.22                 & 8.27 $\pm$ 8.06           & 17.77 $\pm$ 7.73         & 89.13 $\pm$ 17.34 \\
\hline
\end{tabular}
\end{table*}

\section{Results and Discussion}
\label{results}

\subsection{Line Luminosity and Redshift}

We focus on the four strongest forbidden lines present in the spectral range of our composites: \nev~$\lambda3427$, \neiii$~\lambda3870$, \oiii$~\lambda5007$, and \oii~$\lambda3728$.
In addition, we only use the spectra constructed from more than 10 objects.
We trace these four lines across redshift through their luminosities and flux ratios in each of the 7 $L_{5100}$/FWHM(\hb) bins.
In a stratified NLR and with IPs spreading over a range of $\sim 13-100$ eV, we expect the lines with low IP (e.g, \oii; IP =13.5 eV) to be emitted farther away from the SMBH (and thus the source of the ionizing continuum) and therefore from regions with weaker AGN influence than the high IP lines (e.g, \nev; IP =97 eV).

As mentioned in \S \ref{composites}, the selection of objects that went into each of the composites is not dependent on any properties of the NLR.
In addition, by dividing the spectra into bins of $L_{5100}$ and \hb\ FWHM, we aim to ensure that the objects used in making each composite are powered by similar central sources in the sense that the continuum luminosity and the \hb\ FWHM are sensitive to the $L/L_{\rm{Edd}}$ ratios and BH masses of quasars. 
We choose {\em not} to use the derived values of Eddington ratio and $M_{\rm BH}$, but rather the measured values of 5100~\AA\ luminosity and FWHM \hb\ as these are straightforward spectral measurements that are not subject to changes in calibration or uncertain assumptions about bolometric corrections.

\begin{figure*}
\begin{center}
\includegraphics[width= 1.7\columnwidth]{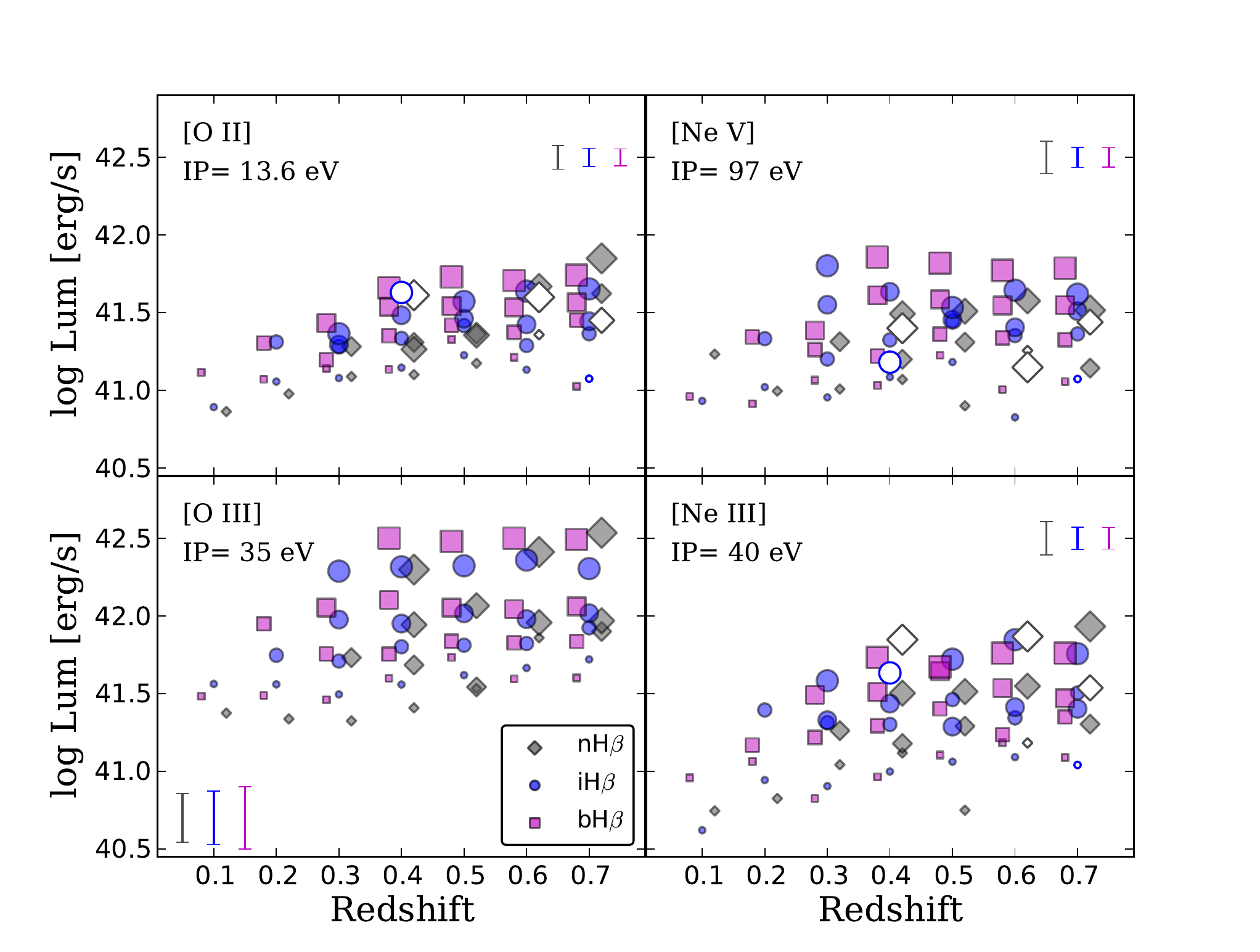}
\caption{Line luminosity vs. redshift for the four narrow lines. Grey diamonds are the composites with \hb\ $<$  2000~ \kms, blue circles are the composites with 2000  $<$ \kms\ \hb\ $<$ 4000~ \kms, and magenta squares are the composites with  4000~\kms\ $<$ \hb  $<$ 20000~ \kms. 
Open symbols denote non-detections. Marker size corresponds to the continuum luminosity bin: larger for higher L$_{5100}$. 
Only composites with $N > 10$ are included. 
The grey/magenta points are slightly offset to the left/right of their redshift value for clarity. 
In all luminosity bins, the \oii\ line is clearly becoming stronger at higher redshifts.
The overall Spearman test gave: $\rho_S({\rm [O~II]})= 0.532$, $p= 0.000$;  $\rho_S({\rm [O~III]})= 0.447$, $p= 0.000$; $\rho_S({\rm [Ne~III]})= 0.479$, $p= 0.000$; $\rho_S({\rm [Ne~V]})= 0.373$, $p= 0.005$ for about 60 objects (detections only).
\label{linelums}}
\end{center}
\end{figure*}

Figure \ref{linelums} shows the line luminosities for the four lines plotted against redshift.
Visually, we see a weak overall tendency of the luminosities of the lower IP lines to increase towards higher redshifts. 
To examine this apparent tend more quantitatively, we calculate the slopes of the line luminosity vs. redshift for each L$_{5100}$ and \hb\ bins along with the correlation coefficients (Table \ref{tbl:linear_reg}).  We discuss the lines individually in the following subsections.

\begin{table*}
\caption{Results of the least-squares regression and the Spearman rank coefficient for the line luminosities and redshift for the four narrow-lines in each of the 7 L$_{5100}$ and \hb\ bins. }
\label{tbl:linear_reg}
\begin{tabular}{llcccc}
\hline
Subset & Line & Slope  & Std Error & $\rho_S$ & $p$ \\ 
\hline
\multirow{4}{*}{\bf n\hb} & \oii & 1.281  & 0.222 &  0.702 & 0.0008\\
				    & \oiii & 1.400  & 0.412 & 0.497 & 0.0171\\
	                             & \neiii & 1.430 & 0.366 & 0.450 & 0.0309\\
	                             & \nev & 0.473  & 0.348 & 0.196 & 0.3811\\
\hline
\multirow{4}{*}{\bf i\hb} & \oii & 0.509 & 0.223 & 0.380 & 0.0625\\
	                            & \oiii & 0.567 & 0.316 & 0.399 & 0.0505\\
	                            & \neiii  & 0.746 & 0.331 & 0.448& 0.0282\\ 
	                            & \nev & 0.494 & 0.322 & 0.120 & 0.5571\\
\hline
\multirow{4}{*}{\bf b\hb} & \oii & 0.550 & 0.249 & 0.543& 0.0056 \\
	                              & \oiii & 0.679 & 0.398 & 0.172& 0.3804\\
	                              & \neiii  & 0.695 & 0.315 & 0.378 & 0.0536\\ 
	                              & \nev & 0.633 & 0.340 & 0.378 & 0.0542 \\
\hline
\multirow{4}{*}{\bf L1} & \oii & 0.297 & 0.114 & 0.510 & 0.026 \\
	                           & \oiii & 0.459 & 0.120 & 0.708 & 0.000\\
	                           & \neiii  & 0.524 & 0.127 & 0.727 & 0.001\\ 
	                           & \nev & 0.014  & 0.147 & 0.088 & 0.729 \\
\hline
\multirow{4}{*}{\bf L4} & \oii & 0.376 & 0.110 & 0.736 & 0.001\\
	                           & \oiii & 0.176 & 0.149 & 0.450 & 0.080\\
	                           & \neiii  & 0.238 & 0.133 & 0.443 & 0.086\\ 
	                           & \nev & 0.033 & 0.121 & 0.209 & 0.438 \\
\hline
\multirow{4}{*}{\bf L7} & \oii & 0.343 & 0.199 & 0.385 & 0.174\\
	                           & \oiii & -0.003 & 0.104 & 0.029 & 0.921\\ 
	                           & \neiii  & 0.109 & 0.188 & 0.188 & 0.519 \\
	                           & \nev & 0.016 & 0.167 & -0.045 & 0.884 \\
\hline
\multirow{4}{*}{\bf L10} & \oii & 0.588 & 0.195 & 0.608 & 0.036 \\
	                             & \oiii & 0.310 & 0.197 & 0.462 & 0.131\\
	                             & \neiii  & 0.464 & 0.183 & 0.626 & 0.029 \\ 
	                             & \nev & -0.451 & 0.297 & -0.607 & 0.083\\
\hline
\end{tabular}
\end{table*}

\subsubsection{L(\oii)}
Compared to the much more prominent \oiii, the \oii\ line is fairly weak in the spectra of radio-quiet quasars.
In general, \oii\ emission may have contributions from photo-ionization from the AGN and/or star-formation in the host galaxy.
Photoionization models suggest that in a pure AGN, the \oii\ to \oiii\ ratio ranges between 0.1 and 0.3 \citep[][and references therein]{ho05}.
Similarly, \citet{kim06} find that in quasars, the AGN contribution to \oii\ is more dominant than the host galaxy's.
The \oii\ luminosities in our composite spectra are shown in top-left panel of Figure \ref{linelums}.
The figure shows a trend of increasing \oii\ luminosity with redshift of nearly an order of magnitude.
We calculated a Spearman correlation coefficient of $\rho_S = 0.54$ for \oii\ luminosity vs. redshift shown in Figure \ref{linelums}, with $p <0.0001$ ($p$ being the probability of no correlation) which indicates the high probability that the two quantities are strongly correlated.
We note however that this correlation can potentially be a result of the missing high luminosity, low-redshift quasar composites in our sample.  
However, tracing the data from quasars of similar luminosity, the general trend is towards higher \oii\ luminosities at higher redshift.  
To further examine this possibility, we look at the correlation between the line luminosity and redshift for each L$_{5100}$ bin separately and we find that the correlation is still strong for most cases but the significance of the correlation becomes much lower (higher $p$ value) as a result of fewer data points in each sub-sample (L1:$\rho_S= 0.510$, L4: $\rho_S= 0.736$, L7: $\rho_S= 0.385$, L10: $\rho_S= 0.608$).
In addition to the Spearman test, we calculate a least-squares regression slope for each L$_{5100}$ and \hb\ bin (Table \ref{tbl:linear_reg}) and find that in most cases, \oii\ has a weak but significant trend to increase towards higher redshift.

\begin{figure}
\begin{center}
\includegraphics[width=1\columnwidth]{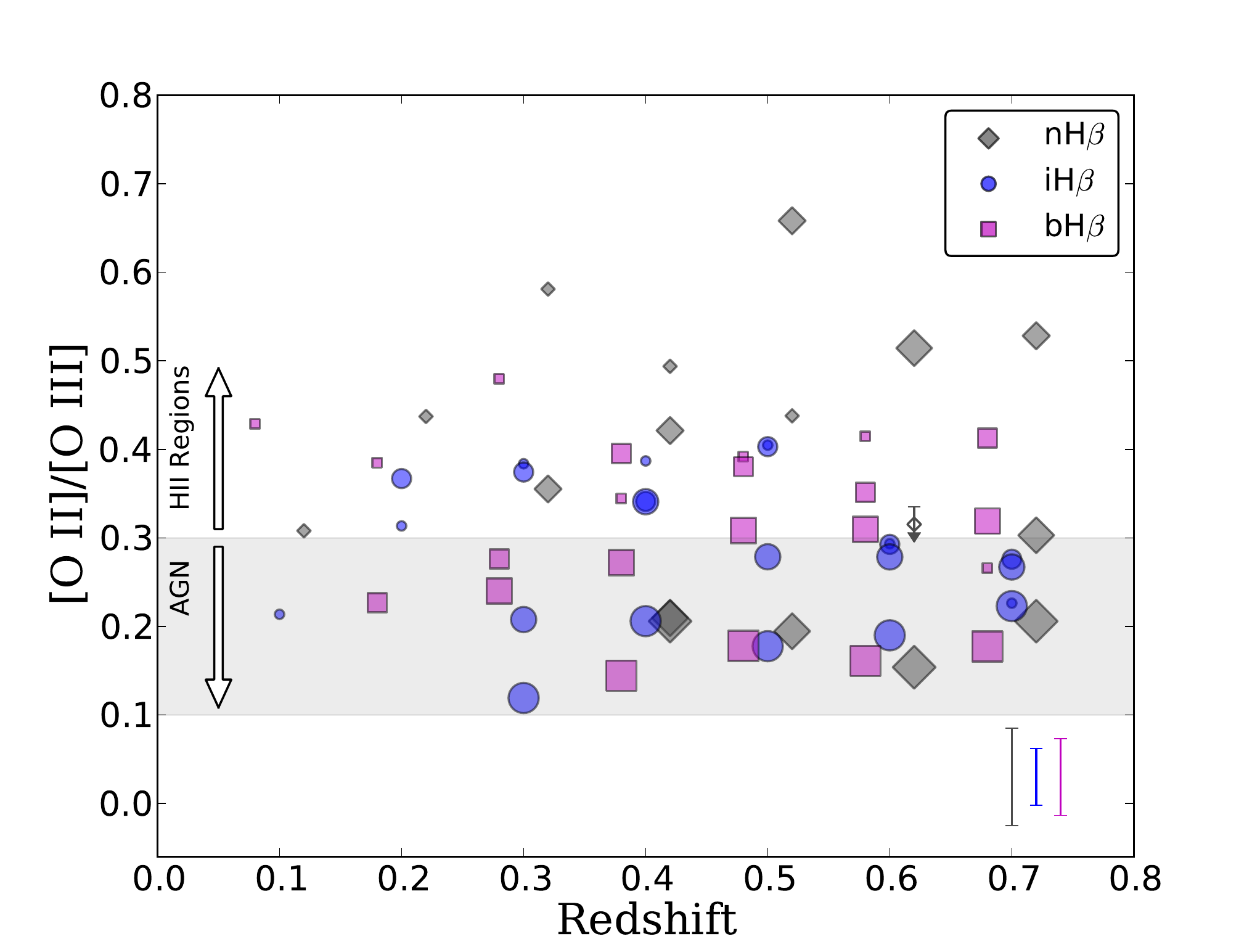} 
\caption{ \oii/\oiii\ flux ratio plotted against redshift. Markers are similar
to Fig. \ref{linelums}.  The shaded region marks the 0.1--0.3 boundary
for an AGN-dominated line ratio; a ratio larger than 0.3 is typical of
stellar H~{\sc ii} regions).  The line ratio distribution shows no
evidence for evolution as a function of redshift.  
\label{oratiovsz}}
\end{center} 
\end{figure}

The \oii\ luminosity is indeed known to be a reliable tracer of SFR in non-AGN and AGN galaxies \citep[e.g.][]{gallagher89, kennicutt98, rosagonz02, hopkins03, kewley04, ho05, kim06,silverman09}.  
This line becomes particularly important for samples with z $\ge 0.5$ as the primary SFR metric of H$\alpha$ has passed out of the optical bandpass \citep[e.g,][]{kewley04}.  
To get a handle on whether star formation might be contributing to the \oii\ emission, we consider the ratio of the line fluxes of the two oxygen lines. 
Figure \ref{oratiovsz} shows the flux ratio of \oii\ to \oiii\ in our sample.  
The figure shows that almost 50\% of the objects (primarily among the low and intermediate continuum luminosity bins; $\log L_{5100} \sim 43 - 44.5$ erg s$^{-1}$) have \oii/\oiii\ $> 0.3$ consistent with \oii\ emission produced in stellar H~{\sc ii} regions \citep{osterbrock06}.  
Therefore, we conclude that there is a contribution from star formation to the \oii\ emission, at least in some objects.

The apparent increase in \oii\ line luminosity as a function of redshift could arise from different factors other than an increase in SFR in the typical quasar host galaxy.  
For instance, the use of a fixed-size aperture (the SDSS uses a 3\arcsec\ fiber) means that spectra potentially include a larger fraction of the host galaxy gas at high redshifts than in lower redshift objects due to the larger projected physical size at higher redshifts.  
Specifically, the projected size is 5.5~kpc at $z = 0.1$ and 22~kpc at $z =0.7$.
However, we do not anticipate this to be the main reason for the observed increase in the \oii\ luminosity (if due solely to photo-ionization by the quasar) as the size of the entire NLR is probably less than 10~kpc. 
To check, we employ the NLR scaling relation from \citet[][their eq. 1]{bennert02}: 
\begin{equation} 
\log R_{NLR} = (0.52 \pm 0.006) \log L_{\rm {[O~ III]}} -(18.5 \pm 2.6),
\end{equation} 
and our measured \oiii\ luminosities (which are significantly higher than the \oii\ luminosities), we find the size of the NLR in our composites is $\sim 4$~kpc. If the \oii\ emission is instead largely from star formation in the host galaxy, then the size of the star-forming region within the galaxy is the relevant factor.  
However, the size of the entire galaxy is not expected to exceed $\sim10$ kpc, and the star-forming regions may be considerably more compact.  
\citet{ichikawa12}, for example, found in a sample of galaxies selected from the MOIRCS Deep Survey with up to redshift 3, that the half-light radius ($R_{50}$) in the $K$-band is between 5--10 kpc.  

\begin{figure}
\begin{center}
 \includegraphics[width=1\columnwidth]{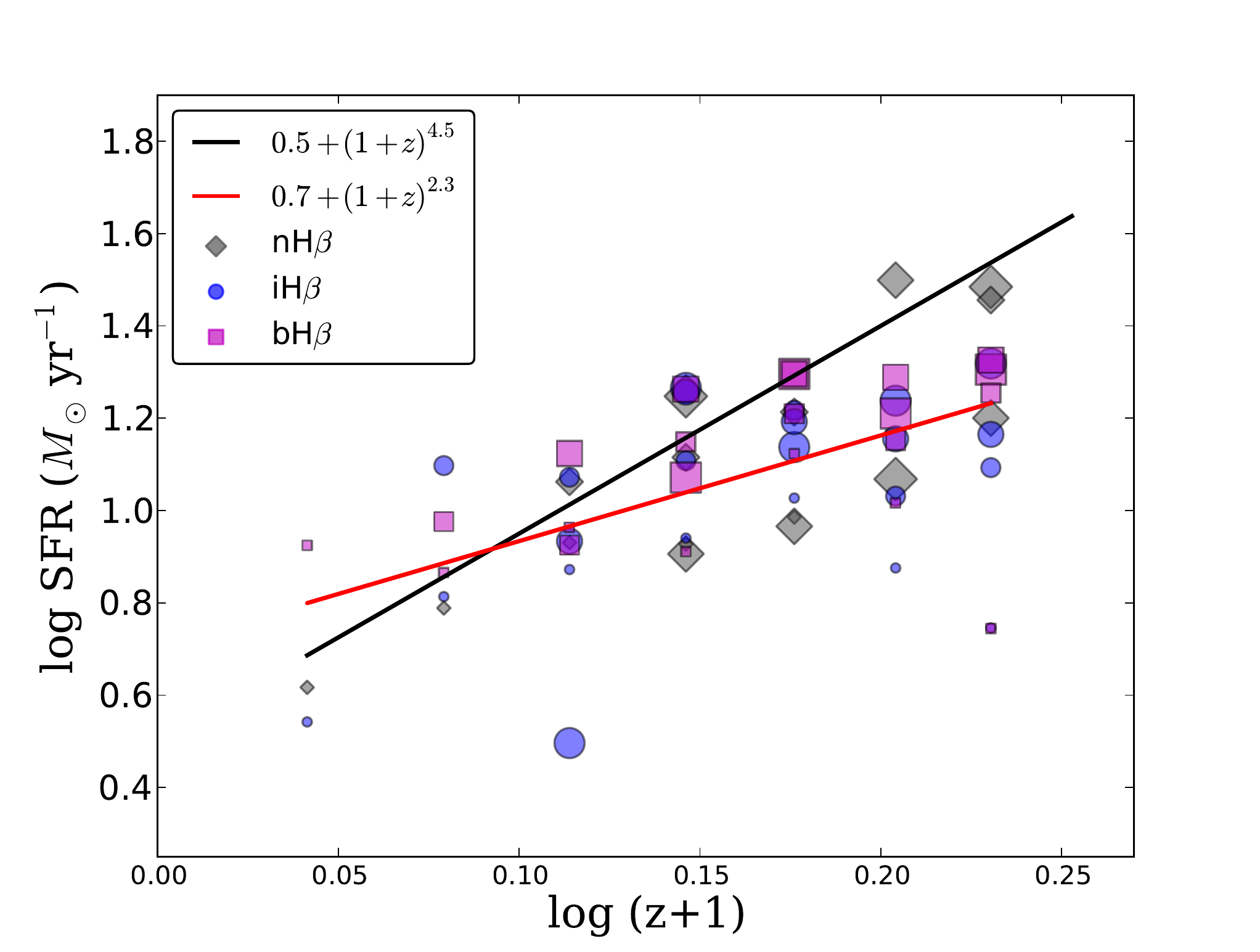}
\caption{SFRs estimated with Eq. \ref{eq:sfr}. 
The points are color-coded to show the narrow H$\beta$ composites in grey, intermediate in blue, and broad in magenta. 
Larger stars indicate higher $L_{5100}$ bins. 
The solid black line is $(1+z)^{4.5}+0.5$ from \citet{rosagonz02} and the red line is our fit with a slope of 2.3$\pm$0.4 and intercept of 0.7. 
The figure shows a clear increase in the estimated SFRs towards higher redshift, though our sample shows less evolution than the non-AGN galaxy sample.  
\label{sfrvsz}}
\end{center}
\end{figure}

The calculation of SFRs using the \oii\ line in isolation is, however, not completely straightforward.  
The conversion of the observed line luminosity to SFRs is complicated by effects such as reddening and metal abundance.  
Several attempts have been made to come up with calibrations that allow for a direct estimation of SFRs from line fluxes \citep[e.g.,][]{rosagonz02, kewley04}.  
These calibrations are useful for estimating SFRs without directly involving assumptions about metallicity and intrinsic extinction that are by and large the major sources of uncertainty.  
We use the L(\oii) calibration equation from \citet{rosagonz02} for non-AGN galaxies to estimate SFRs for our composite spectra: 
\begin{equation} 
SFR_{\rm{[O~II]}} = 8.4 \times 10^{-41} L_{\rm{[O~II]}}~ \rm{M_{\odot} yr^{-1}}, 
\label{eq:sfr}
\end{equation} 
where we have first subtracted 10\% of the \oiii\ luminosity from the \oii\ luminosity to account for the AGN contribution, i.e., we have assumed all of the \oiii\ emission is from the AGN, and assumed an \oii\ to \oiii\ line ratio of 0.1.  
We plot the results in Figure \ref{sfrvsz} which shows that the SFR calculated with equation \ref{eq:sfr} increases with $z$.  
As interpreted at face value, the figure also shows that higher continuum luminosity composites have higher SFRs as expected from the higher \oii\ luminosities in the higher $L_{5100}$ composites (see Fig. \ref{linelums}).  
The different \hb\ FWHM bins overlap in terms of SFRs.  
We are also interested in comparing the slope of SFR vs. $z$ for our sample and a sample of non-AGN galaxies.  
\citet{rosagonz02} found a best fit for the z = 0--1 range of $SFR \propto$ (1+z)$^{4.5}$; the line representing this fit is shown in Fig. \ref{sfrvsz} (with an offset of 0.5 to match our data).  
Our best fit is also shown in the figure with a slope of $2.3\pm0.4$ which is notably smaller than the \citet{rosagonz02} fit.  
Given the assumptions that have gone into converting our \oii\ luminosities (which have an uncertain contribution from the quasar) into SFRs, it is not clear that this indicates a significant discrepancy between the samples.

\subsubsection{The AGN Lines}
\label{otherlines}
With higher IPs, the \oiii, \neiii, and \nev\ lines can be thought of as AGN lines (IPs: 35, 40, and 97 eV, respectively).
After accounting for the lack of high luminosity, low redshifts sources, the luminosities of \neiii\ and \oiii\  are almost constant with redshift as the two lower panels of Figure \ref{linelums} show.
With almost equal IPs of 35 and 40 eV, \oiii\ and \neiii\ are both expected to respond similarly to the ionizing radiation and consequently have similar behaviour across redshift.

We calculate Spearman correlation coefficients for each line vs. redshift and find: $\rho_S(\rm{[O~III]}$) = 0.447 ($p <$ 0.0001), $\rho_S(\rm{[Ne~III]}$) = 0.479 ($p <$ 0.0001).
The slight increase in the luminosities of these two lines might be a result of the contribution of the host galaxy ISM which undergoes enhanced SFRs at higher redshift.  
However, a comparison of the apparent increase in \oiii\ luminosity with redshift to the error bars (lower left corner of Fig.~\ref{linelums}), and tracing the trend within each luminosity bin indicates to us that this apparent correlation is not real.

As for the \nev\ line (Fig. \ref{linelums}), the correlation is  still significant but is much weaker with $\rho_S(\rm{[Ne~V]}$) = 0.373 ($p$ = 0.005).
Table \ref{tbl:linear_reg} lists all the slopes and the Spearman correlation coefficients calculated for the luminosity and \hb\ bins separately for the line luminosity vs. redshift correlations.

\begin{figure}
\begin{center}
\includegraphics[width=1\columnwidth]{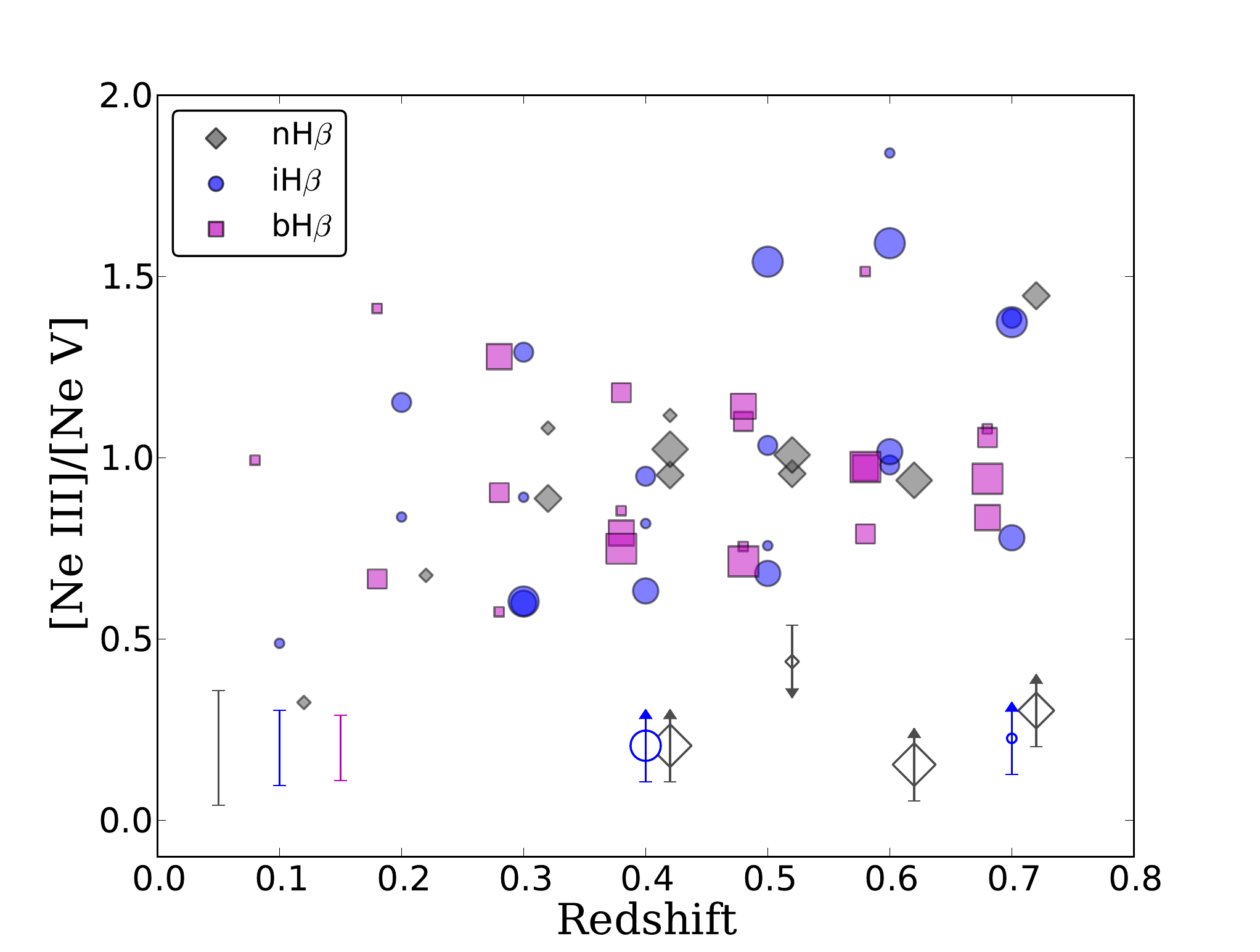}
\caption{\neiii/\nev flux ratio. The ratio does not show a clear trend with redshift indicating no evidence for a change in the ionizing spectrum with redshift.
\label{neratiovsz}}
\end{center}
\end{figure}

This again supports the notion of a pure AGN line with luminosity that is not changing up to redshift 0.75.
The behaviour of \nev\ is as expected in a scenario where the ionizing radiation (and consequently the ionization parameter) is not changing over redshift.
The \neiii/\nev\ flux ratio shown in Fig. \ref{neratiovsz} does indeed show evidence for no change in the hardness of the ionizing radiation in the redshift range of 0--0.75.

\subsection{\oiii, \feii\ and Eigenvector 1}
\begin{figure}
\begin{center}
\includegraphics[width=1\columnwidth]{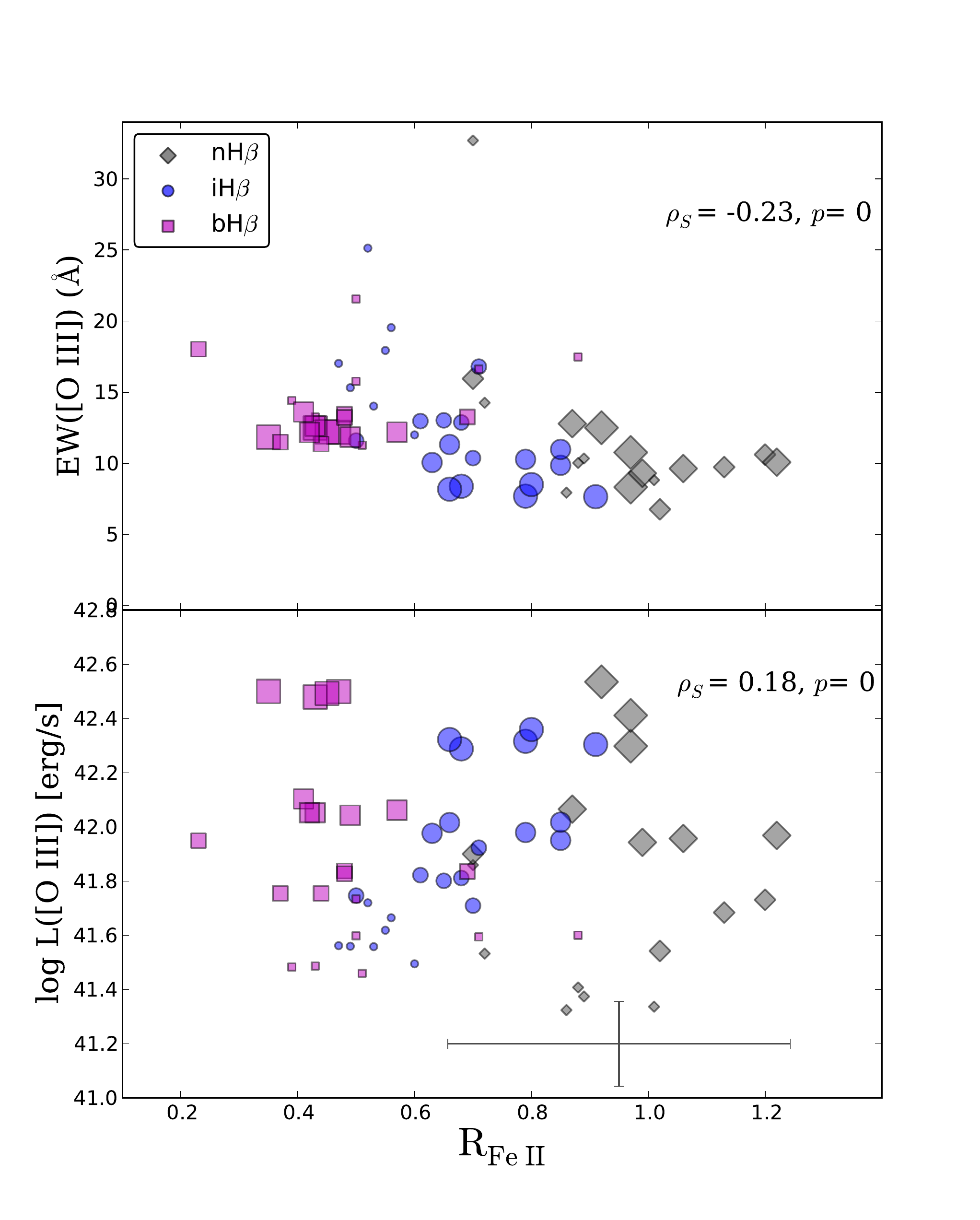}
\caption{Top: The \oiii\ EW is plotted against the R$_{\rm \feii}$ ratio.
The marker shapes correspond to \hb\ width (as labeled in the legend) and the sizes are proportional to the $L_{5100}$ bin.
The two quantities are strongly anti-correlated with $\rho_S =-0.63$.
Bottom: \oiii\ luminosity plotted against R$_{\rm \feii}$. 
Average error bars are shown.
The two quantities appear to be uncorrelated with $\rho_S= 0.18$.
\label{o3vsfe}}
\end{center}
\end{figure}

A suite of correlations among quasar spectral properties have been identified, most notably from Principal Component Analysis (PCA) \citep[e.g,][]{boroson92, boroson02, sulentic07}. 
One strong PCA correlation that is particularly related to the spectral range we are focusing on is referred to as Eigenvector 1 \citep[EV1;][]{boroson92,boroson02}. 
EV1 describes an inverse correlation between the strengths of \oiii$\lambda5007$ and the \feii$\lambda4570$ feature (the strength of \feii\ is often quantified as the EW ratio of \feii\ to \hb, defined as $\rm{R_{FeII}}$).
In addition to \oiii\ and \feii, EV1 also includes correlations among quantities such as \hb\ width, line blueshift and asymmetry, and the radio-loudness \citep[e.g,][]{bachev04}.
This is usually seen as objects with strong \feii\ have weaker \oiii, narrower, symmetric \hb, and are more likely to be radio-quiet.
In Fig. \ref{o3vsfe}, we plot the \oiii\ luminosity and EW vs. $\rm{R_{FeII}}$.
The $\rm{R_{FeII}}$ values in this figure are derived from the median values of the \hb\ and \feii\ from the \citet{shen11} measurements.
Figure \ref{o3vsfe} shows that composites with narrower \hb\ do indeed have stronger \feii.
Our composite spectra (see Appendix \ref{compositelib} in the online version) also show that objects with narrower \hb\ have stronger, more conspicuous \feii.
Figure \ref{o3vsfe} (top panel) also shows that the \oiii\ EW and the $\rm{R_{FeII}}$ ratio are anti-correlated --in agreement with the EV1 findings.
We also examine the correlation between the \oiii\ and \feii\ using the \oiii\ line luminosity (Fig. \ref{o3vsfe}, lower panel) but we find no correlation between the two quantities.
We note that the original EV1 studies \citep{boroson92, boroson02} used the equivalent width of \oiii\ (or its ratio to the \hb\ line) to quantify its strength rather than the luminosity.
The \oiii\ and the \feii\ lines are emitted in two regions with different physical properties (e.g, density and ionization parameter) and at significantly different size scales (i.e., kpcs vs. pcs).
Despite being repeatedly found, the presence of a correlation between these two lines has not been fully understood.
The predominant factor that appears to be driving this correlation is the accretion rate \citep{boroson92, boroson02, shen14}.
In this picture, objects at high accretion rates have thicker accretion disks which blocks radiation from reaching the narrow-line region leading to weaker \oiii\ emission, however, this scenario does not account for the correlation that include \feii.

\section{Summary and Conclusion}
In this work we examine the narrow forbidden lines in quasar optical spectra across redshift and up to z = 0.75. 
To enhance the S/N, we utilize a large sample of quasars from the SDSS-DR7 and create 62 median composite spectra (with at least 10 objects in each composite).
We indirectly constrain the properties of the central source (accretion rate and BH mass) by grouping objects with similar $L_{5100}$ and H$\beta$ values in bins that we use to make the composites.

Our results can be summarized as follows:
\begin{enumerate}

\item The luminosity of the lowest ionization narrow line,\oii, appears to increase with redshift (top-left panel of Fig. \ref{linelums}).
Because the \oii\ line is known to be a reliable SFR estimator in non-active galaxies, we interpret this result as evidence for higher SFRs in the host galaxies of AGNs at higher redshift, 
We calculate a SFR using the \oii\ luminosity and the scaling relation of \citet{rosagonz02}  and find that the estimated SFR is increasing with redshift however with a smaller power-law slope ($2.3\pm0.4$) than found in non-AGN galaxies (4.5; \citealt{rosagonz02}.

\item The strength of the correlation between the line luminosity and redshift appears to depend on the IP of the line.
The luminosity of the line with the lowest IP (\oii) has the strongest correlation with redshift while the ``intermediate'' IP lines (\neiii\ and \oiii) have weaker dependancies and the \nev\ with IP =97 eV appears to be independent of redshift (Fig. \ref{linelums}, see also Table \ref{tbl:linear_reg}).
This behaviour is in line with the idea of a stratified NLR where lines with higher IPs are emitted closer to the central source (which is similar for $z<0.75$) with negligible contribution from the host galaxy gas while lines with low IPs are emitted farther out and also have more contribution from stellar sources of ionizing radiation.  

\item We find that the strength of the broad \feii\ emission is
anti-correlated with both the \oiii\ EW and the width of \hb, i.e., objects with weaker \oiii\ (lower EWs) tend to have narrower \hb\ and stronger \feii\ (larger $\rm{R_{FeII}}$, see Fig. \ref{o3vsfe}).  
This result is in agreement with the definition of EV1 \citep{boroson92, boroson02}.  
The presence of a correlation between these lines has been attributed to the accretion rate of the central source (L/L$_{Edd}$).  

\end{enumerate}

The study of the impact of AGNs on their host galaxies allows us to probe galaxy evolution and the possible role that SMBHs play in it.
AGN narrow-line regions are the zones where the interaction between the ionizing radiation from the accretion disk and the host galaxy ISM takes place and can therefore serve as excellent probes of the AGN--host galaxy interface.  
The increase in the SFRs at higher redshift (up to z $\sim$ 1--2) is well-known in non-AGN galaxies and appears to show similar behaviour in AGN hosts.  
Examining higher redshifts to explore this trend further will require near-IR spectroscopy.

\section*{acknowledgments}

We are grateful to to Fran\c{c}oise Combes and Niel Brandt for thoughtful discussions.  
This work was supported by the Schlumberger Foundation - Faculty for the Future Program (A.T.), the Natural Science and Engineering Research Council of Canada, and the Ontario Early Researcher Award Program (A.T., S. C. G.).  
This research made use of Astropy, a community-developed core Python package for Astronomy \citep{astropy13}.  
Funding for the SDSS and SDSS-II has been provided by the Alfred P. Sloan Foundation, the Participating Institutions, the National Science Foundation, the U.S. Department of Energy, the National Aeronautics and Space Administration, the Japanese Monbukagakusho, the Max Planck Society, and the Higher Education Funding Council for England. 
The SDSS Web Site is http://www.sdss.org/.  
We thank the anonymous referee for helpful comments that improved this paper.


\appendix
\section{Quasar Spectral Library}
\label{compositelib}

This spectral library consists of 62 quasar spectra generated from a homogeneous set of quasars that share similar continuum luminosity, broad line width, and are at similar redshifts. 
Starting from a sample of $\sim 16,000$ quasars, we separate the objects into 4 groups of continuum luminosity (high, two intermediate, and low) and further separate each of these luminosity groups into 3 groups depending on their H$\beta$ width (using 2000 and 4000 ${\rm km s^{-1}}$ as boundaries).  
We further bin these groups at redshift steps of 0.1, 0.2, 0.3, 0.4, 0.5, 0.6, and 0.7.  
{\bf The results are shown below. A full coloured version of the figures as well as the FITS files are available online}.

\begin{figure*}
\label{z1}
\includegraphics[width=1.5\columnwidth]{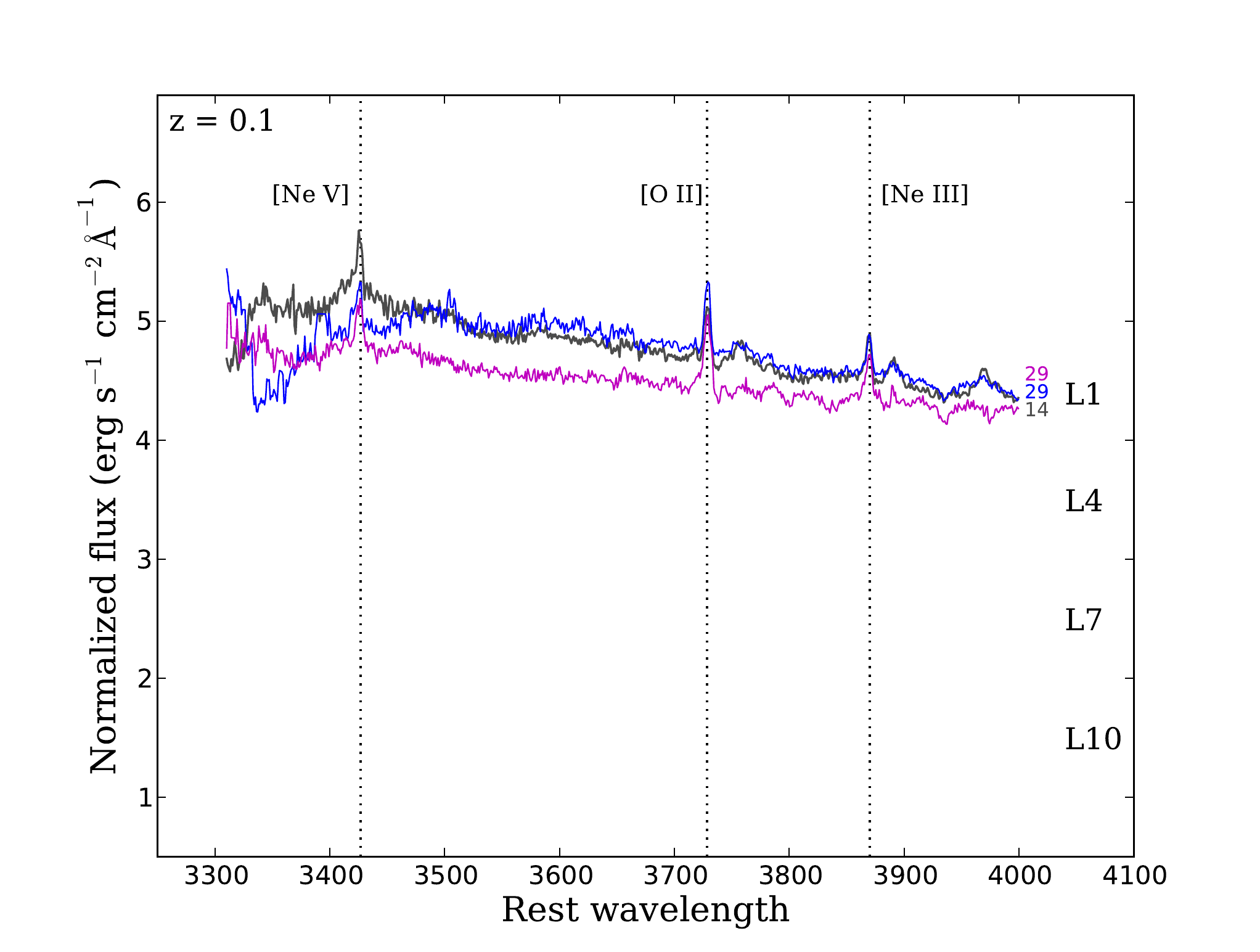}
\caption{Median composite spectra at redshift z=0.1. Top is the high continuum luminosity (L10) and bottom is the low luminosity (L1). Spectra at the same luminosity level but with different widths of H$\beta$ are overplotted with narrow H$\beta$ in grey, intermediate H$\beta$ in blue, and broad H$\beta$ in magenta.  To the right side, we show the number of individual spectrum used in making each composite. Only composites with $> 10$ objects are shown. L1, L4, L7 and L10 indicate the continuum luminosity subsets with L1 being the lowest and L10 the highest. At z = 0.1, only the lowest luminosity bin has data.}
\end{figure*}

\begin{figure*}
\label{o3z1}
\includegraphics[width=1.5\columnwidth]{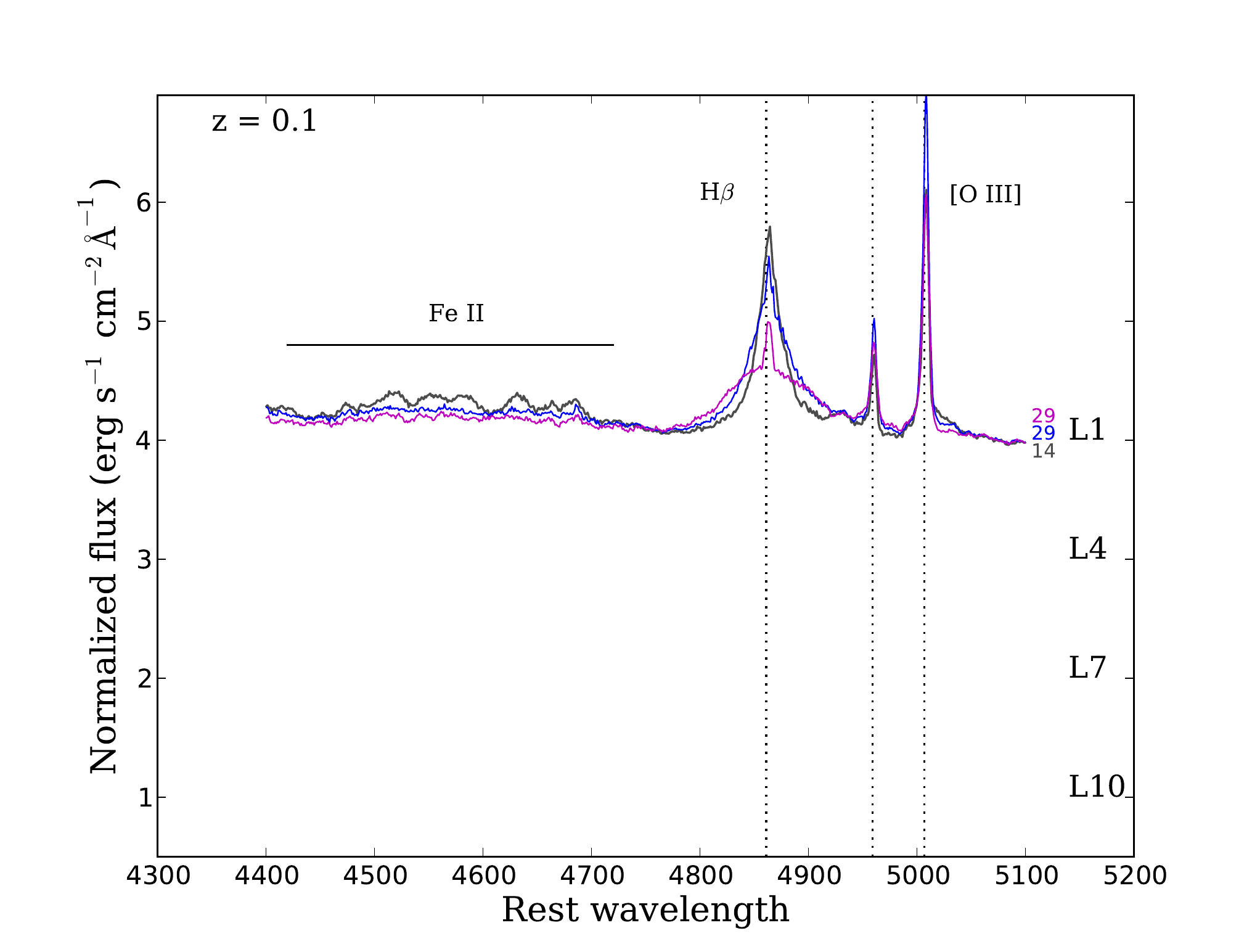}
\caption{Similar to Fig. \ref{z1} but for the wavelength range covering H$\beta$ and [O~{\sc III}].}
\end{figure*}

\begin{figure*}
\includegraphics[width=1.5\columnwidth]{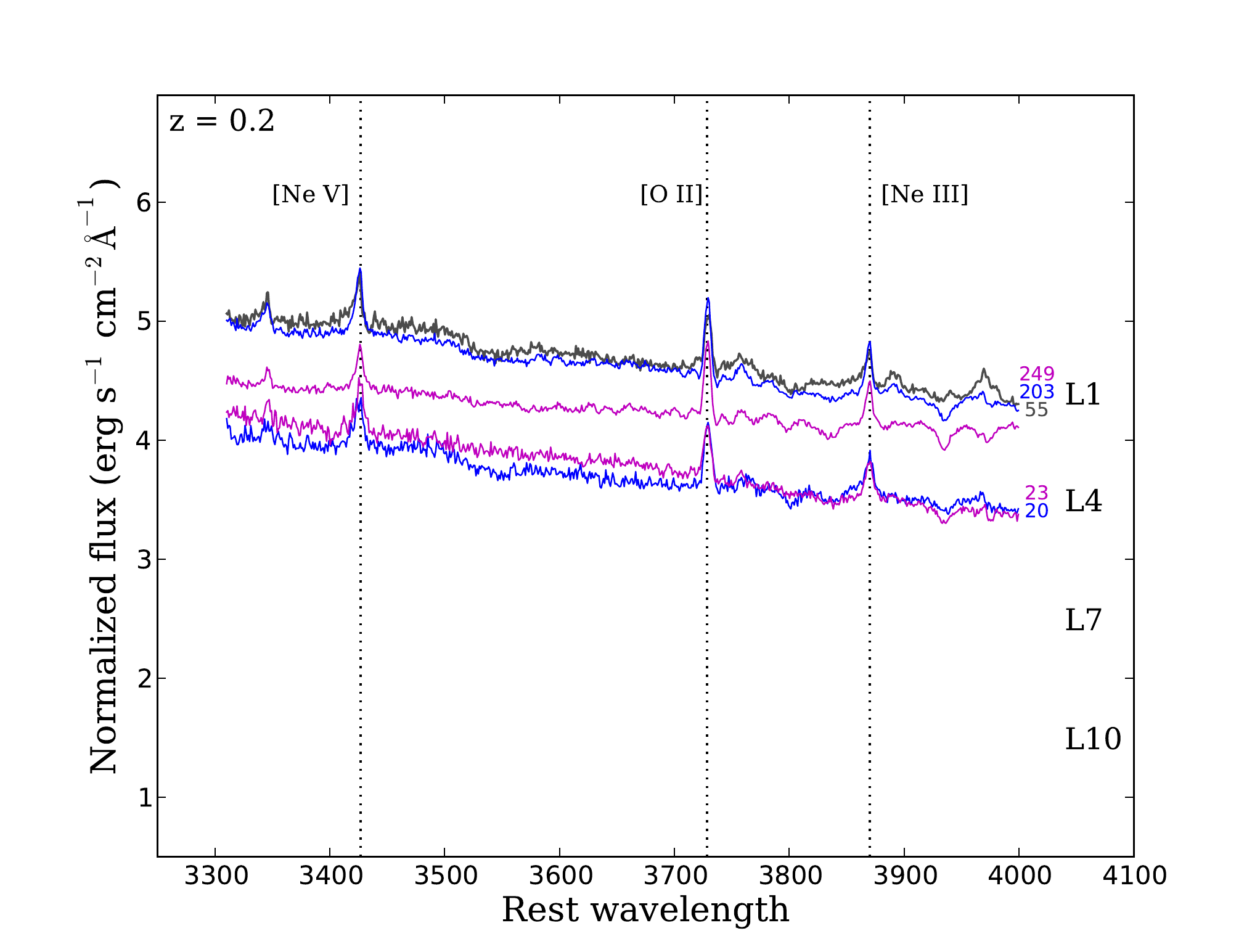}
\caption{Same as Fig. \ref{z1} but for redshift bin centred at z= 0.2.}
\end{figure*}

\begin{figure*}
\includegraphics[width=1.5\columnwidth]{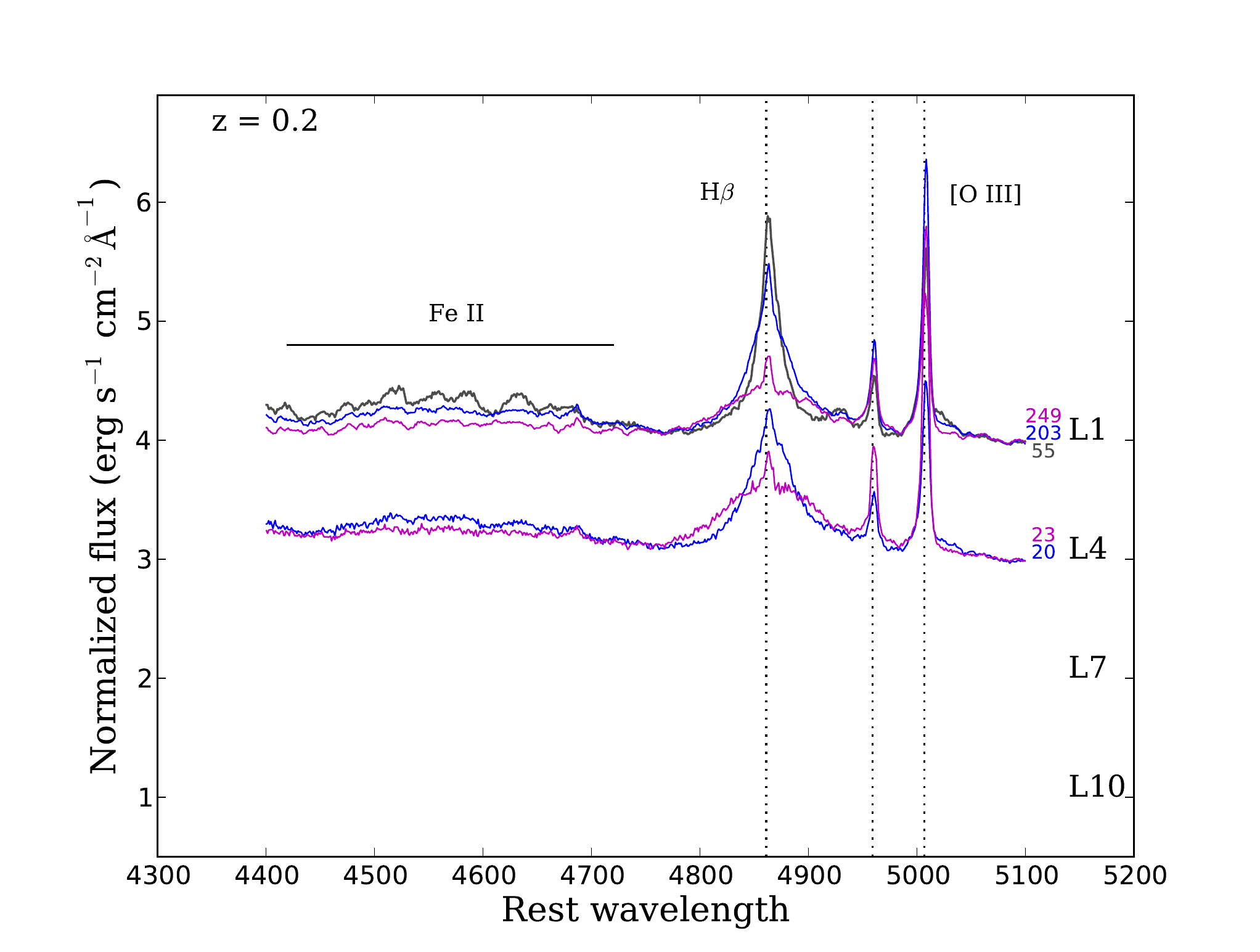}
\caption{Same as Fig. \ref{o3z1} but for redshift bin centred at z= 0.2.}
\end{figure*}

\begin{figure*}
\includegraphics[width=1.5\columnwidth]{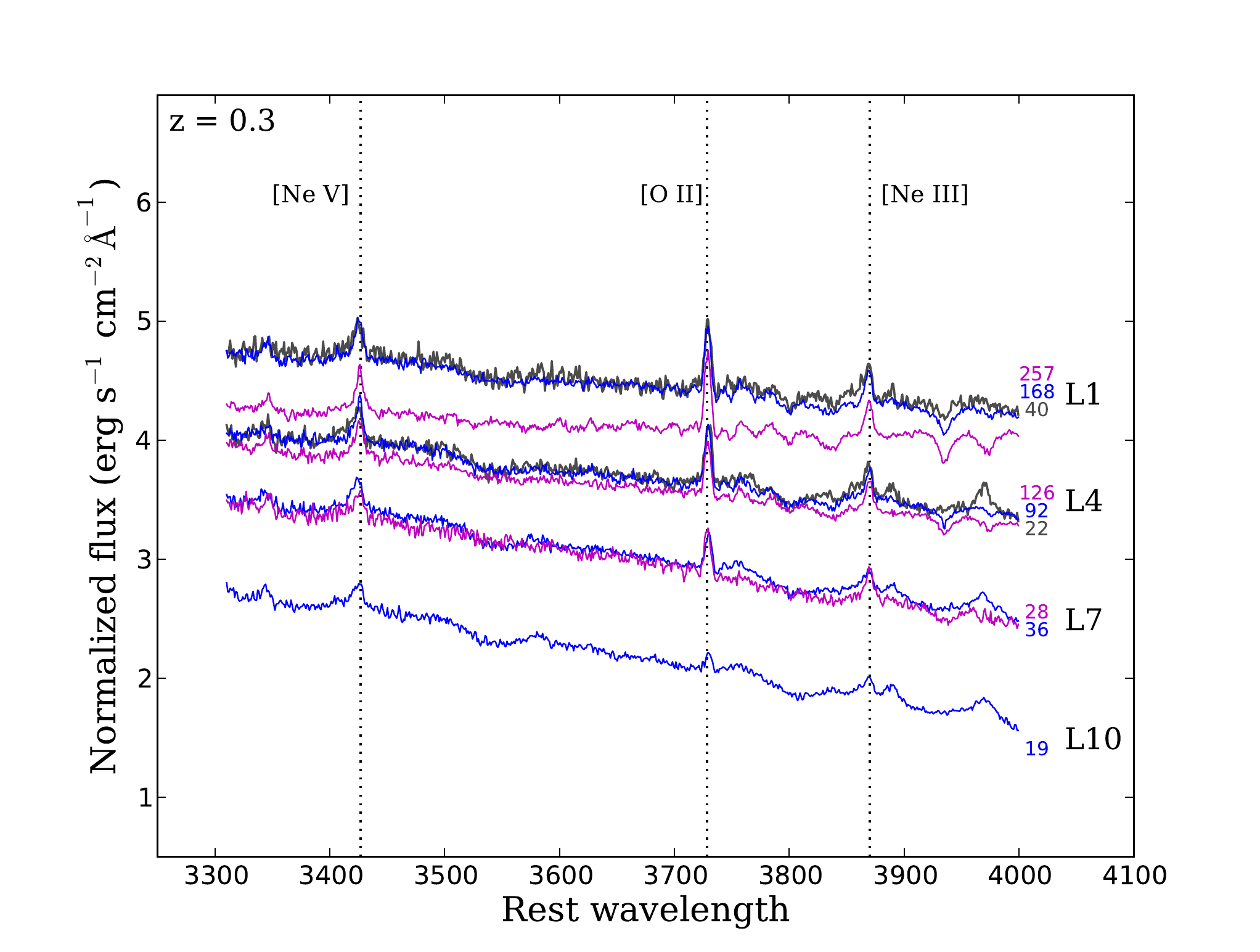}
\caption{Same as Fig. \ref{z1} but for redshift bin centred at z= 0.3.}
\end{figure*}

\begin{figure*}
\includegraphics[width=1.5\columnwidth]{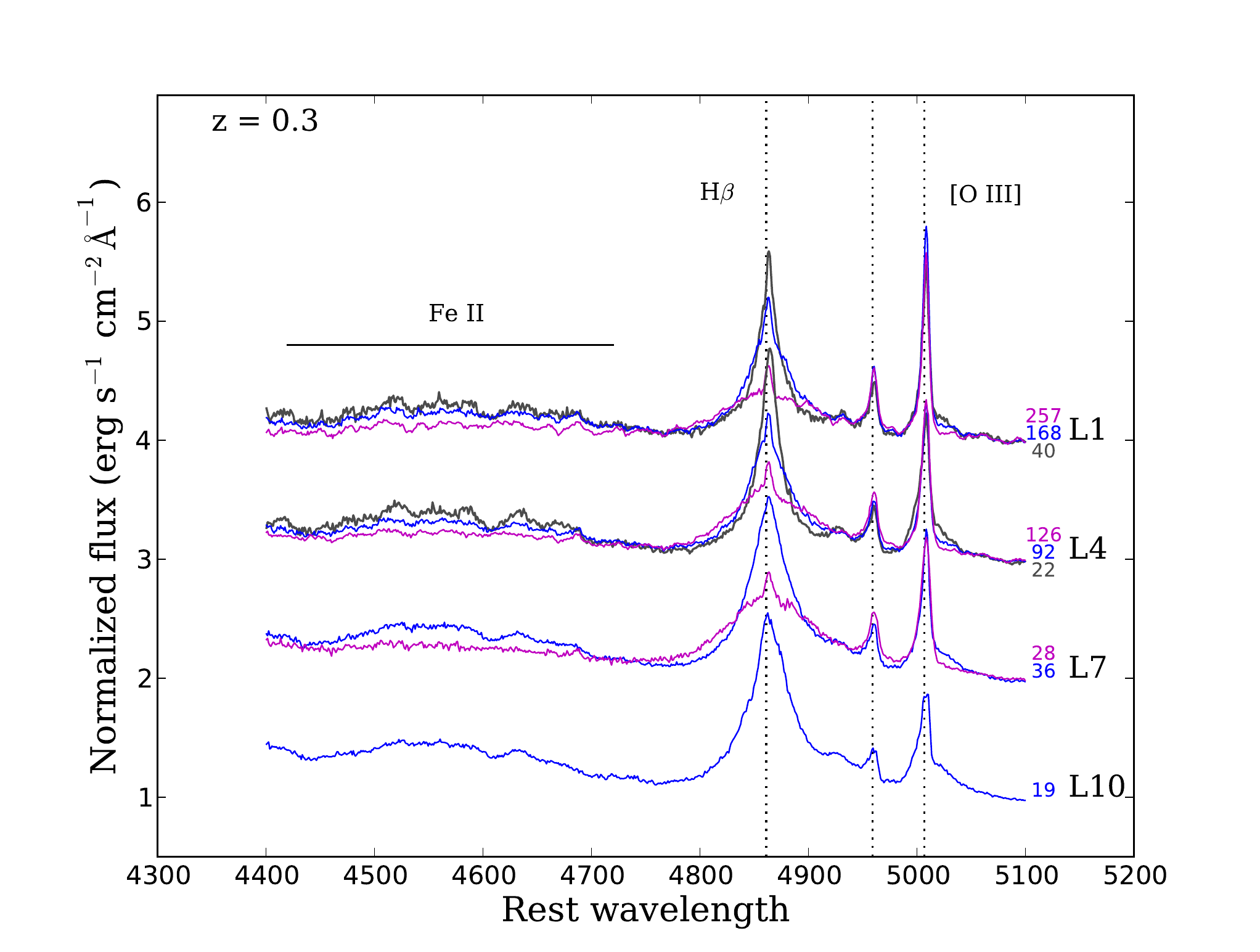}
\caption{Same as Fig. \ref{o3z1} but for redshift bin centred at z= 0.3.}
\end{figure*}

\begin{figure*}
\includegraphics[width=1.5\columnwidth]{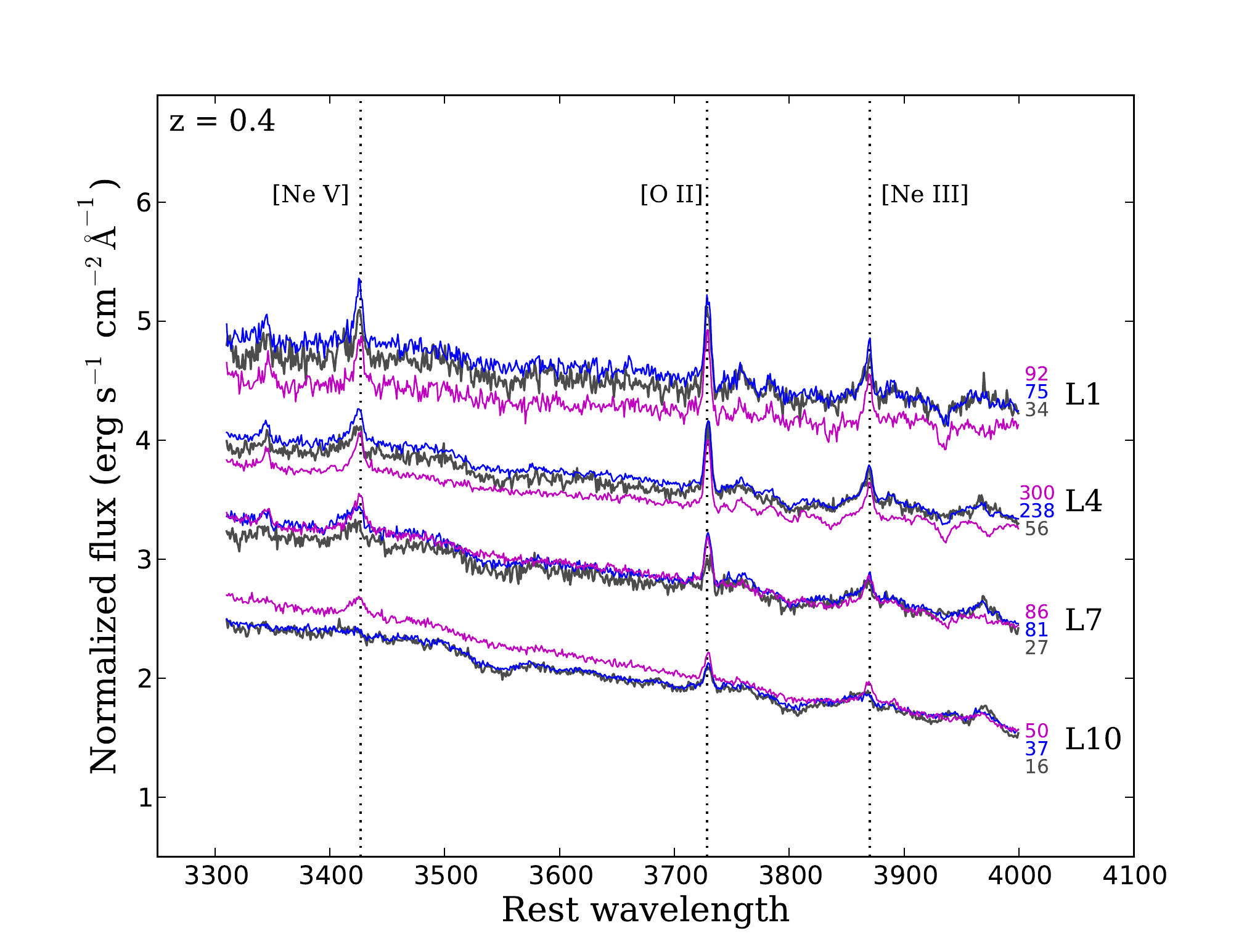}
\caption{Same as Fig. \ref{z1} but for redshift bin centred at z= 0.4.}
\end{figure*}

\begin{figure*}
\includegraphics[width=1.5\columnwidth]{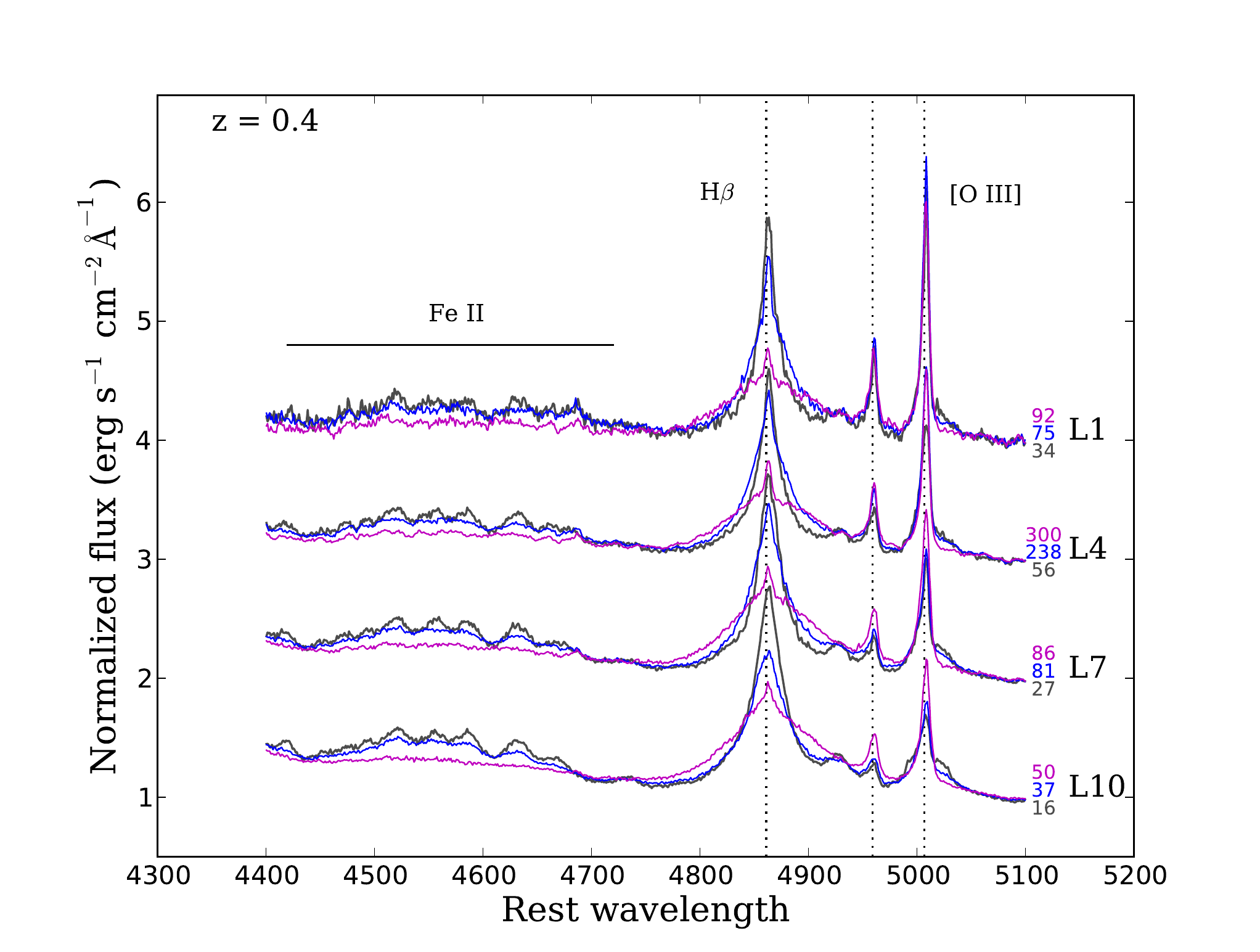}
\caption{Same as Fig. \ref{o3z1} but for redshift bin centred at z= 0.4.}
\end{figure*}

\begin{figure*}
\includegraphics[width=1.5\columnwidth]{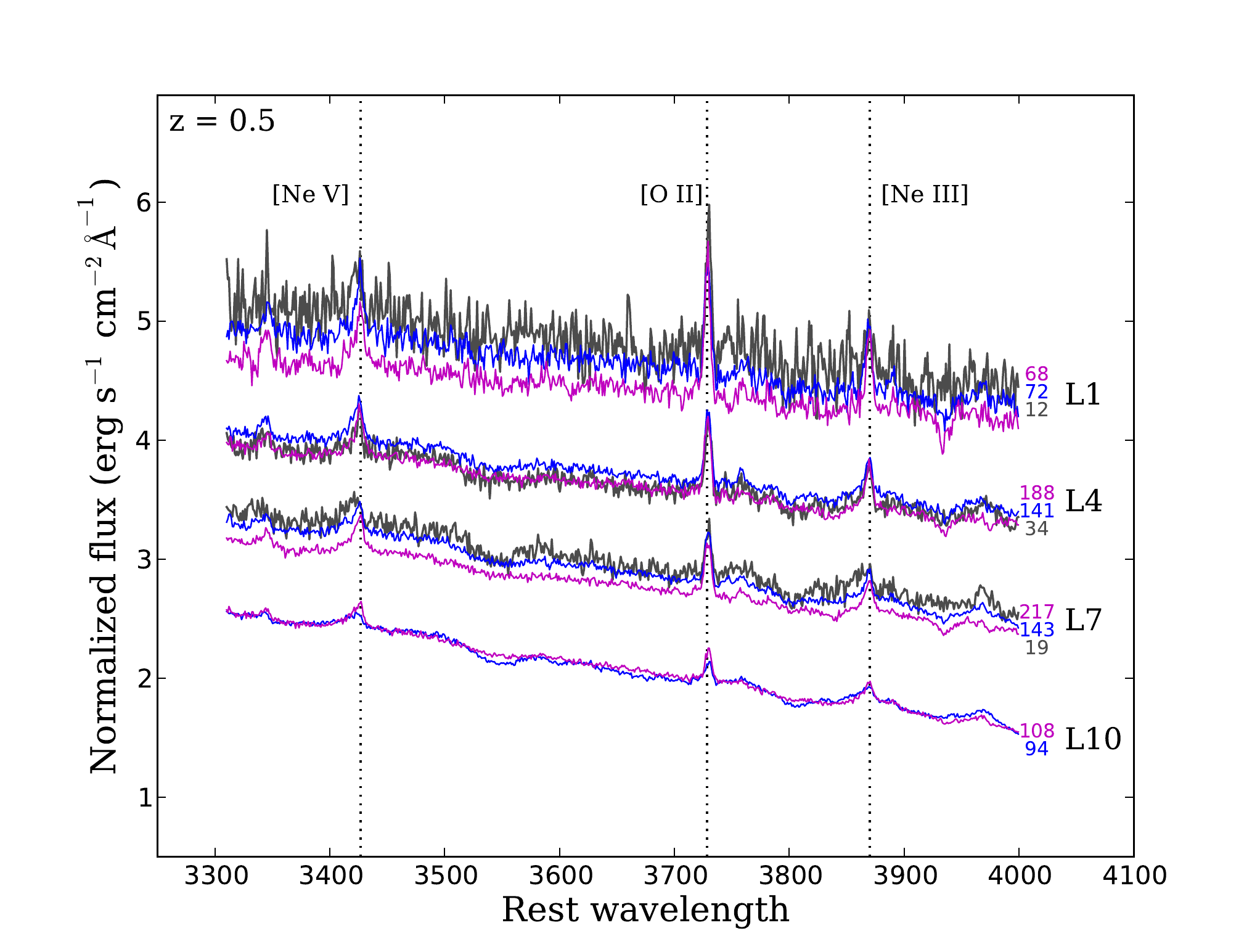}
\caption{Same as Fig. \ref{z1} but for redshift bin centred at z= 0.5.}
\end{figure*}

\begin{figure*}
\includegraphics[width=1.5\columnwidth]{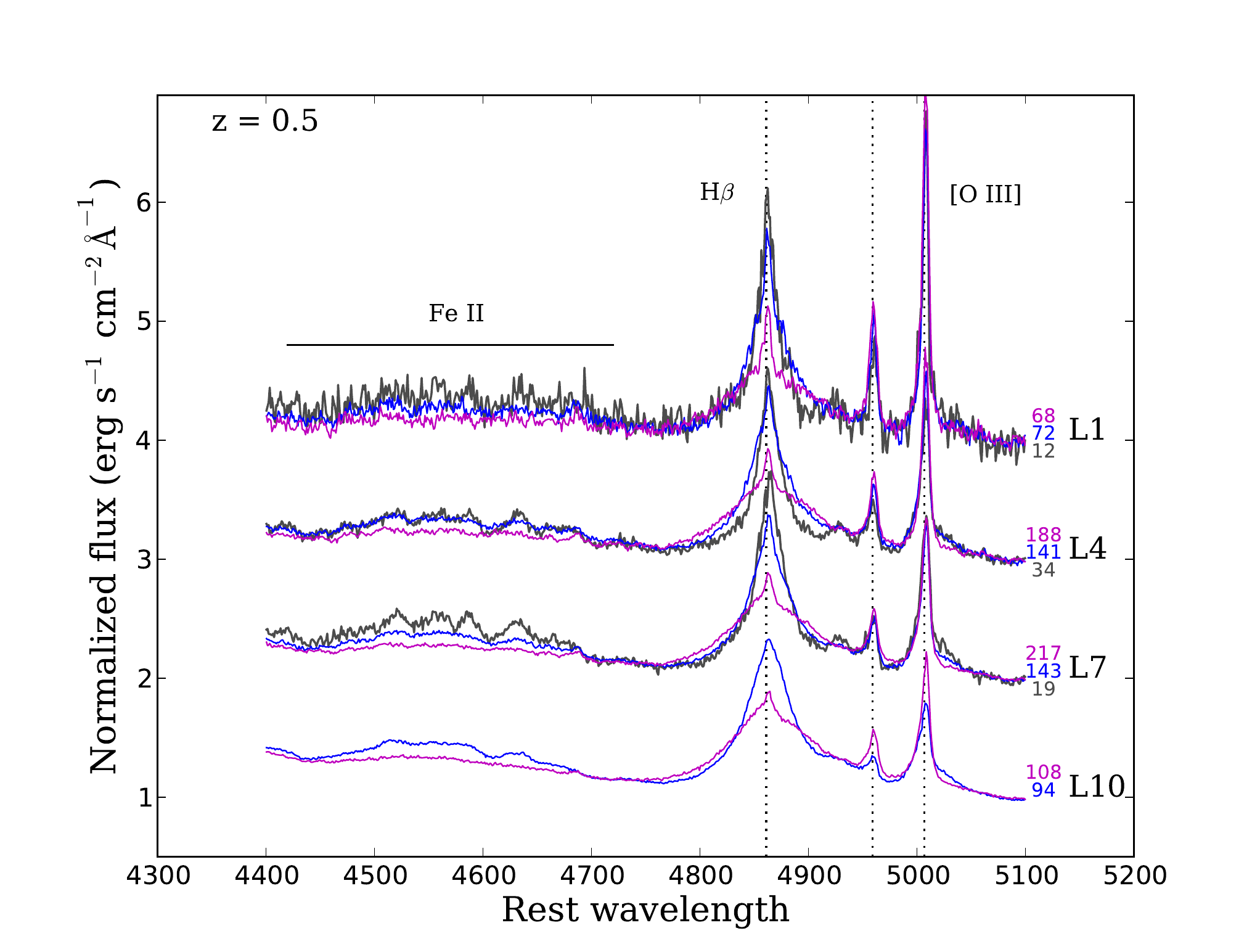}
\caption{Same as Fig. \ref{o3z1} but for redshift bin centred at z= 0.5.}
\end{figure*}

\begin{figure*}
\includegraphics[width=1.5\columnwidth]{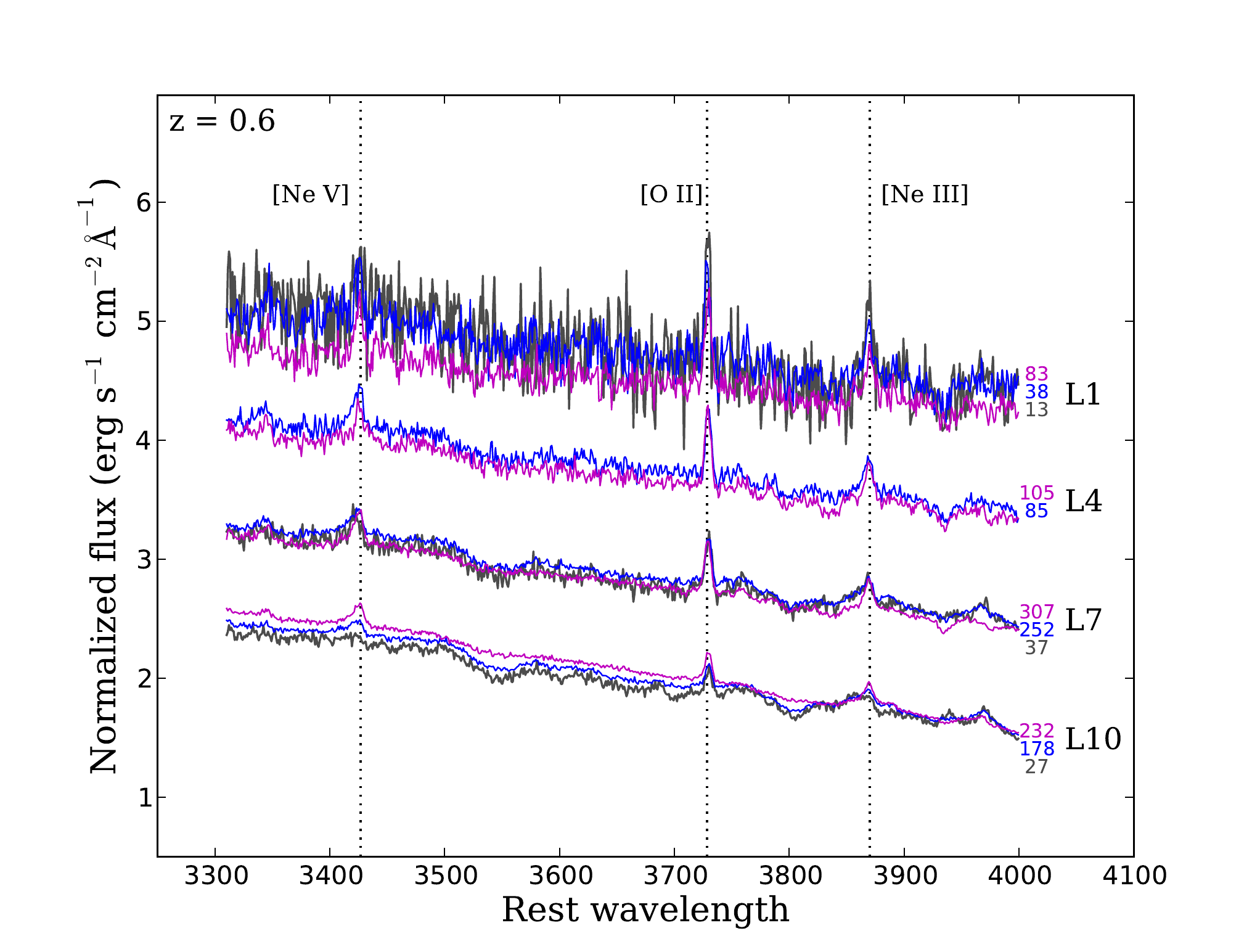}
\caption{Same as Fig. \ref{z1} but for redshift bin centred at z= 0.6.}
\end{figure*}

\begin{figure*}
\includegraphics[width=1.5\columnwidth]{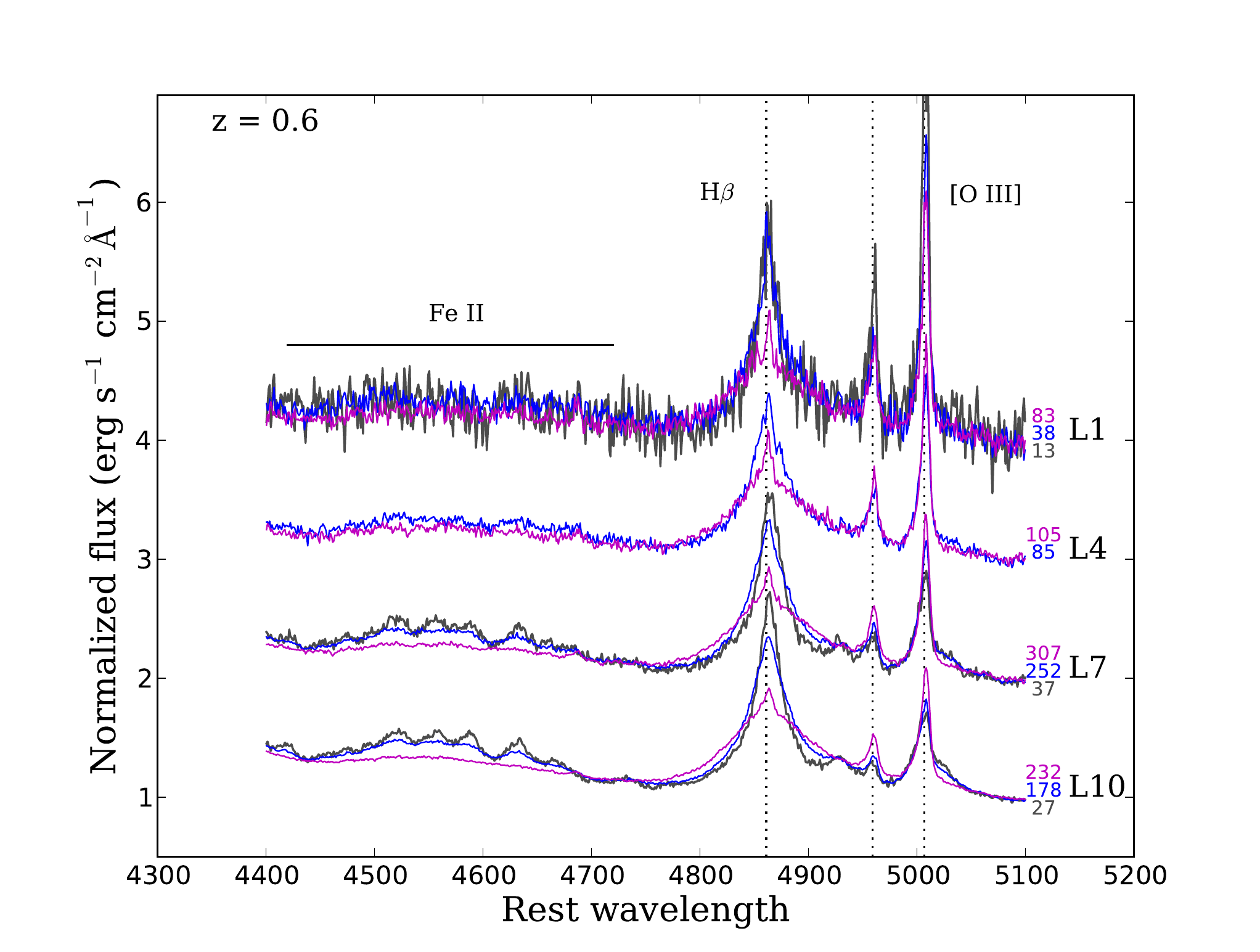}
\caption{Same as Fig. \ref{o3z1} but for redshift bin centred at z= 0.6.}
\end{figure*}

\begin{figure*}
\includegraphics[width=1.5\columnwidth]{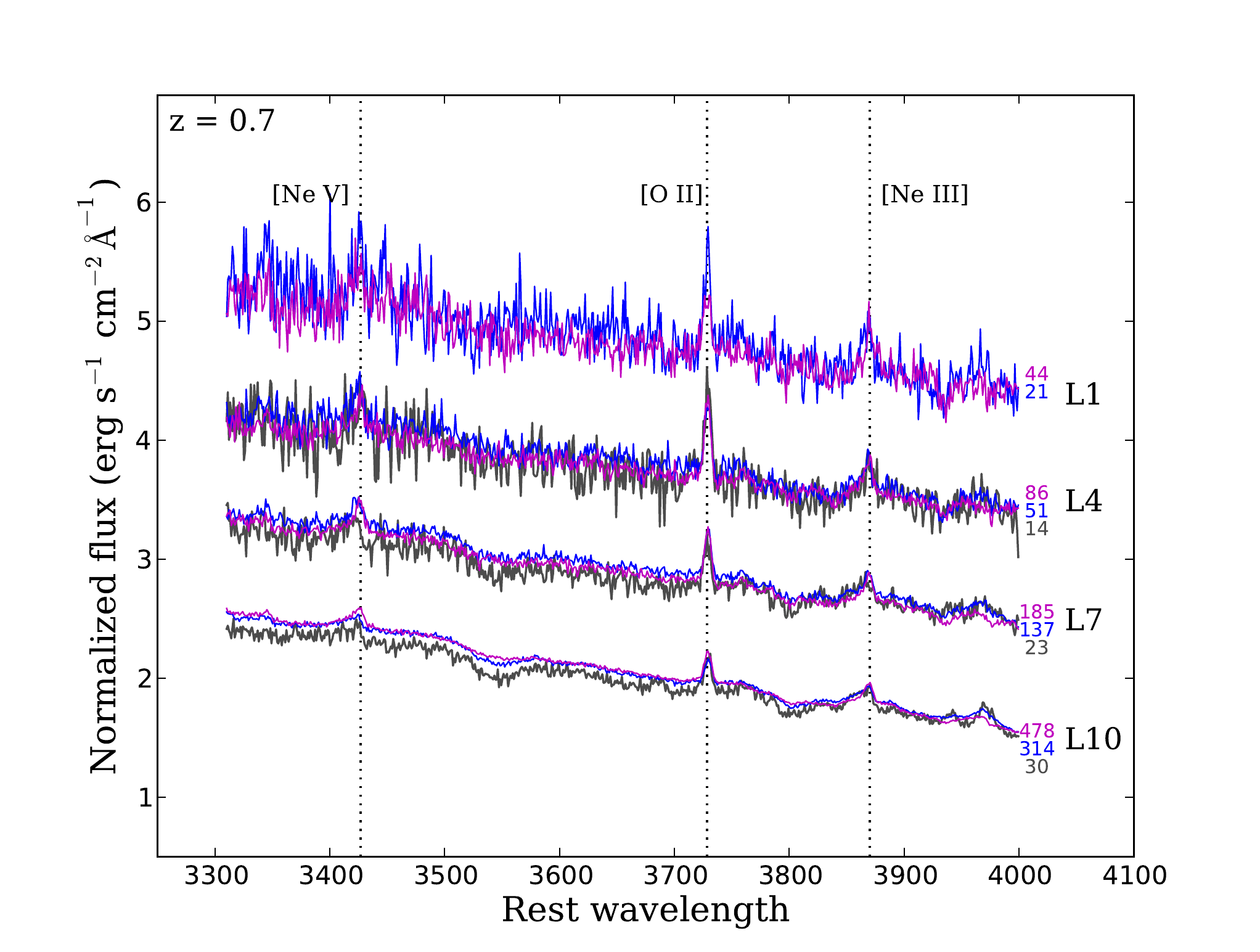}
\caption{Same as Fig. \ref{z1} but for redshift bin centred at z= 0.7.}
\end{figure*}

\begin{figure*}
\includegraphics[width=1.5\columnwidth]{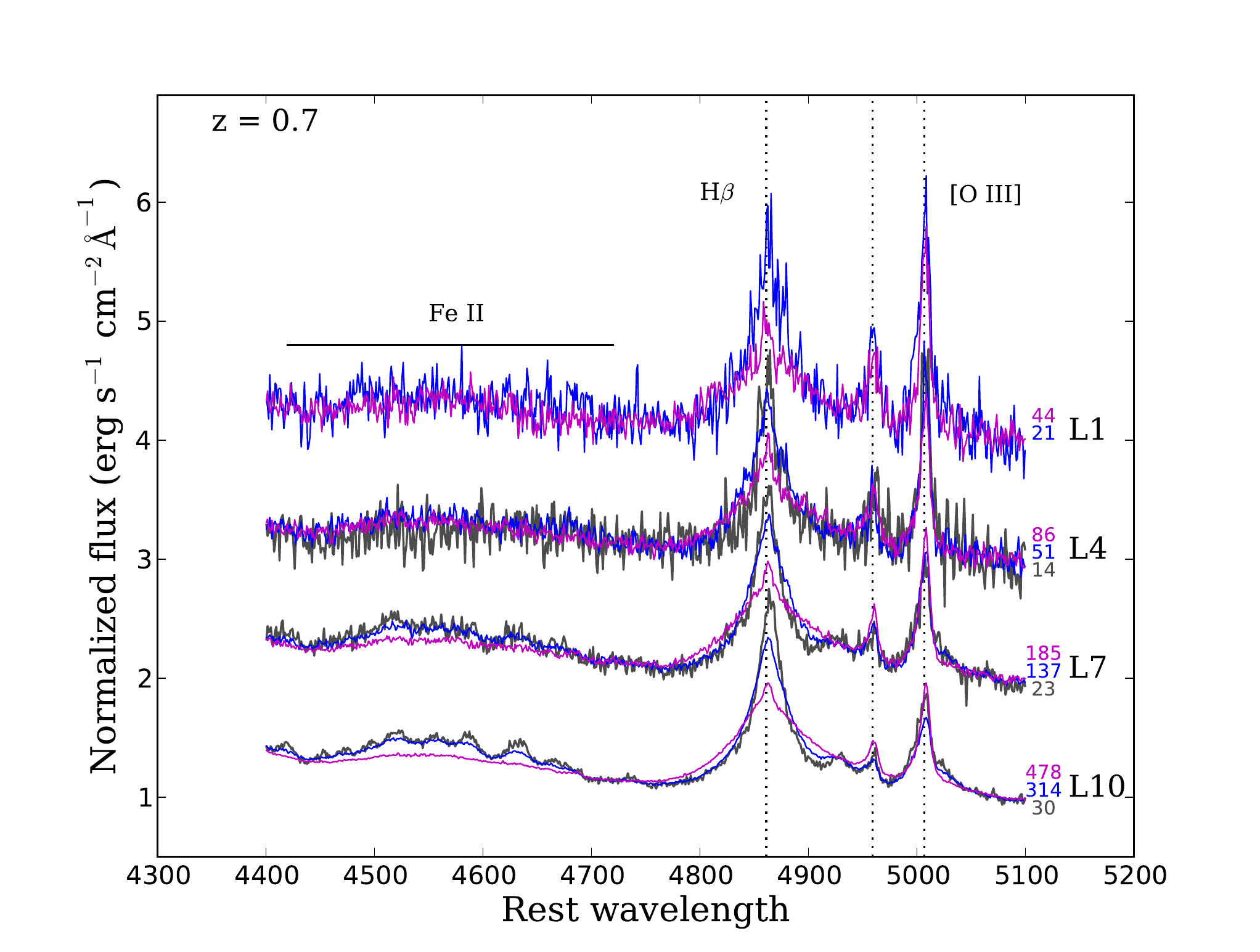}
\caption{Same as Fig. \ref{o3z1} but for redshift bin centred at z= 0.7.}
\end{figure*}

\end{document}